\documentclass[journal]{IEEEtran}
\ifCLASSINFOpdf
\else
\fi
\usepackage{cite}
\usepackage{graphicx} % for pdf, bitmapped graphics files
\usepackage{amsmath}
\usepackage{mathtools}
\usepackage{amssymb}
\usepackage{xcolor}
\usepackage{booktabs, tabularx}
\usepackage{fixdif}
\usepackage{empheq}
\usepackage{comment}
\usepackage{ragged2e}
\usepackage{bm}
\usepackage{url}
\usepackage{hyperref}
\usepackage[normalem]{ulem} % enables \sout

\begin{document}
\title{Stiffness-Aware Decentralized Dynamic State Estimation for Inverter-Dominated Power Systems}

\author{Xingyu~Zhao,~\IEEEmembership{Student~Member,~IEEE,~}Marcos~Netto,~\IEEEmembership{Senior~Member,~IEEE,~}and~Junbo~Zhao,~\IEEEmembership{Senior~Member,~IEEE}

\thanks{This work is partially supported by the U.S. National Science Foundation under grants 2328241 and 2442160 (Corresponding author: Junbo Zhao).}% <-this % stops a space
\thanks{\textcolor{black}{X. Zhao and J. Zhao are with the Thayer School of Engineering, Dartmouth College, Hanover, NH 03755 USA. (e-mail: junbo.zhao@dartmouth.edu)}}% <-this % stops a space
\thanks{M. Netto is with the Department of Electrical and Computer Engineering, New Jersey Institute of Technology, Newark, NJ 07102 USA.}% <-this % stops a space

}
\markboth{}%
{Shell \MakeLowercase{\textit{et al.}}: Bare Demo of IEEEtran.cls for IEEE Journals}
\maketitle

\begin{abstract}
Dynamic state estimation (DSE) is becoming increasingly important for monitoring inverter-dominated power systems. Due to their cascading control structures, inverter-based resources (IBRs) exhibit multi-timescale dynamics, leading to stiff system models that pose significant challenges for conventional DSE methods. In particular, explicit discretization schemes often require impractically small sampling intervals to maintain numerical stability, increasing computational and communication burdens.%\sout{Modern power systems are increasingly dominated by inverter-based resources (IBRs). Due to their cascading control structures, IBRs exhibit multi-timescale dynamics, leading to stiff dynamic models that pose new challenges for dynamic state estimation (DSE). Existing DSE methods may require very small sampling intervals to ensure numerical stability, thereby increasing the computational and communication burden.} 
To address this issue, this paper proposes a stiffness-aware decentralized DSE method for inverter-dominated power systems. The statistical linearization is used to construct a local linear surrogate model for the nonlinear dynamics, which allows matrix-exponential discretization to enable analytical uncertainty propagation in discrete time, rather than relying on explicit integration schemes. This enables stable DSE at lower sampling rates. Numerical results reveal the mechanism by which stiff dynamics destabilize conventional DSE and demonstrate that the proposed method achieves efficient and accurate estimation under coarse sampling conditions.
\end{abstract}

% Note that keywords are not normally used for peer-reviewed papers.
\IEEEpeerreviewmaketitle

\section{Introduction}
\IEEEPARstart{T}{he} modern power system is undergoing a transformative shift from a synchronous-machine-dominated system to a power-electronics-inverter-dominated one. Functioning as the interface of distributed energy resources, inverters synchronize with the grid through digital controllers, making their dynamic states defined by their control. For system operators, real-time visibility into these dynamic states is of particular interest, as it enables better interpretation of system dynamics and faster response to abnormal conditions. For instance, several large solar farms were incorrectly tripped in California's 2016 Blue Cut Fire event due to erroneous phase-locked loop (PLL) behaviors \cite{NERC:PLL}. With DSE, system operators may identify early signs of PLL loss of synchronism and take proactive actions before instability develops. Compared with port measurements alone, hidden dynamic states can provide more informative evidence for assessing IBR operating modes, such as whether an IBR is subject to current limiting \cite{fan2022currentlimit}, entering momentary cessation \cite{Savastianov:MC}, or successfully riding through a fault. Moreover, DSE can be integrated with the dissipating energy flow method to identify the root cause of resonance and oscillatory behavior \cite{marchi2023dse}, which benefits system operators by enabling both real-time oscillation source location and post-fault analysis. Beyond providing state information, DSE also offers a statistical approach for anomaly detection through innovation and residual tests. This can help system operators detect abnormal conditions, including model parameter mismatches, cyberattacks \cite{Bamigbade:cyber}, and sensor malfunctions \cite{zhao2018robustdecentralized}. Finally, DSE can also be extended to online parameter estimation. By augmenting the state vector, parameters, such as virtual inertia and damping, can be jointly estimated and tracked in real time \cite{tan2024adaptive}. For a comprehensive review of DSE applications, we refer to \cite{Zhao:role, Liu:protection}.

Generally, DSE is formulated within a Bayesian filtering framework, which consists of two iterative steps: a prediction step that computes the prior state estimate, and a correction step that updates the posterior state estimate. However, unlike synchronous machines, IBRs introduce tightly coupled cascaded control loops operating across microsecond-to-second timescales. This results in severely stiff dynamic models, whose eigenvalue spectra can span several orders of magnitude. Such stiffness fundamentally challenges conventional DSE, which typically relies on explicit discretization schemes designed for moderately stiff synchronous-machine dynamics. %{\sout{Due to the cascaded control structures of IBRs, their models are very \emph{stiff}.}}
As the prediction step requires an explicit discrete-time state-transition function, the stiffness of the dynamic model poses a major challenge to maintaining numerical stability in current DSE methods. This paper aims to address this problem.

\subsection{Literature Review}
\justifying
\subsubsection{Discretization} Explicit methods, such as the forward Euler and Runge--Kutta methods have been adopted in DSE \cite{Qi:Euler, Tan:RK}, although their numerical stability is guaranteed only under limited conditions \cite{Lara:revisit}. When applied to stiff equations, explicit methods require impractically small time steps to keep the error in the result bounded. For DSE, using tiny time steps may be constrained by either computational or communication costs or by the sampling rate of measurement systems. To address this issue, implicit methods such as backward Euler \cite{Katanic:BackEuler} and trapezoidal \cite{Qiu:trapez} can be leveraged. However, implicit methods involve nested calculations, which are more computationally demanding and lack direct control over computational burden. For example, when applying implicit methods to an unscented Kalman filter (UKF) or a particle filter (PF), a set of equations needs to be solved numerically multiple times at each step size. Moreover, solving such equations may require evaluating the Jacobian of the dynamic model, which is difficult to obtain for IBRs with proprietary or highly complex control structures. In \cite{Xingyu:GFM}, a discrete-time model with numerical stability guarantee is derived for grid-forming inverters (GFMs). However, this is still far from a generic solution, since it relies heavily on the assumption of time-scale separation between control loops and the linear structure of each subsystem, which are satisfied by GFMs only. From the modeling perspective, a commonly used practice is to combine model order reduction with explicit methods. However, the use of reduced-order models in DSE can introduce significant bias at the onset of transients \cite{Zhao:Robust}. Moreover, the reduced-order model may be truncated to different dimensions under different setpoints, further complicating DSE design. To date, a numerically stable and computationally efficient DSE method for the stiff IBR models is still lacking.

\subsubsection{DSE for IBRs} Existing research regarding DSE for IBRs mainly adopts the decentralized scheme \cite{Singh2014}. In decentralized DSE schemes, port voltage measurements are typically treated as system inputs for driving the component dynamics, while port current measurements serve as output measurements for state correction. Early-stage studies on decentralized DSE for IBRs focus on those that operate in grid-following mode, such as solar photovoltaic systems \cite{Zhang:PV} and wind generators \cite{Yu:DFIG, Zhu:DFIG, Ma:DFIG, Huang:DSE}. Recently, DSE has also been extended to IBRs operating in grid-forming mode \cite{Zhao:Robust, Huang:DSE, Zhao:GFM}, as well as to hybrid IBRs that switch between grid-forming and grid-following controls \cite{Huang:DSE2, odunlami2025hybrid}. However, existing studies either use reduced models that overlook fast transients caused by inner-loop controllers and PLLs (\cite{Zhao:Robust, Yu:DFIG, Zhu:DFIG, Ma:DFIG, Zhao:GFM}), or discretize the IBR model through explicit methods with extremely small time steps (\cite{Zhang:PV, Huang:DSE, Huang:DSE2}). The challenges associated with small time steps have been identified in \cite{Zhu:DFIG} and addressed through interpolation and step-size control. However, the method in \cite{Zhu:DFIG} still relies on the explicit Runge--Kutta scheme, whose limitations have already been discussed. To date, enabling decentralized DSE with detailed IBR models under relaxed sampling rates remains an open problem. 

\subsection{Contributions} 
Although extensive efforts have been devoted to extending DSE formulation to handle the nonlinearity and uncertainty of power system dynamics, the challenge of numerical stability in DSE has remained largely overlooked, despite the inherent stiffness of power system models. The major contributions of this paper are summarized as follows:
\begin{itemize}
\item 
This paper identifies stiffness-induced numerical instability as a critical yet largely overlooked issue in decentralized DSE. It shows that, when detailed IBR models with multi-timescale dynamics are used for DSE, a conventional prediction step based on explicit methods may require impractically small sampling intervals to maintain numerical stability.

\item 
This paper proposes a stiffness-aware decentralized DSE method. It constructs a local continuous-time surrogate of the original nonlinear system using statistical linearization and derives analytical uncertainty propagation using matrix-exponential discretization. Meanwhile, the statistical linearization error is explicitly quantified and incorporated into the covariance prediction. By avoiding direct explicit discretization of the original stiff nonlinear system, it improves numerical stability under coarse sampling conditions. Moreover, its Jacobian-free formulation makes it suitable for detailed IBR models with complex or partially inaccessible control structures.

\item 
This paper supports the proposed method with theoretical analysis and numerical studies. Theoretical analysis shows that the one-step prediction errors of the state mean and covariance are second order in the sampling interval. Numerical results reveal that, under coarse sampling, the proposed method improves numerical robustness and estimation accuracy over explicit-discretization-based DSE methods, while incurring much lower computational cost than implicit-discretization-based alternatives.
\end{itemize}

This paper proceeds as follows: Section \ref{proposed} introduces the proposed stiffness-aware decentralized DSE method and analyzes its one-step prediction error. Section \ref{numerical} presents numerical comparisons of the proposed method against both the explicit and the implicit-discretization-based UKF on inverter-dominated power system test cases.

\section{Proposed Method}
\label{proposed}
This section proposes a modified UKF scheme for decentralized DSE, in which the conventional prediction step is replaced by a stiffness-aware,
Jacobian-free scheme based on statistical linearization and matrix exponential discretization. 
\subsection{Problem Formulation}
A dynamic component (e.g., SM or IBR) in a power system can be described by the following state-space equations
\begin{subequations}
\allowdisplaybreaks
\label{eq:ssa}
\begin{align}
 \frac{\d \bm x(t) }{\d t} &= \bm f\left(\bm x (t) , \bm u(t) \right), \label{eq:dyn}\\
 \bm z(t) &= \bm h(\bm x(t), \bm u(t)) 
\end{align}
\end{subequations}
where $\bm x \in \mathbb{R}^n$ denotes the vector of dynamic states (e.g., rotor angle and frequency), $\bm u \in \mathbb{R}^m$ denotes the input vector, and $\bm z \in \mathbb{R}^p$ denotes the output vector. Both the input and output variables correspond to the port variables (e.g., terminal voltage and current) that couple the component dynamics with the external power network. The functions $\bm f:\mathbb{R}^{n}\times\mathbb{R}^{m}\rightarrow\mathbb{R}^{n}$ and $\bm h:\mathbb{R}^{n}\times\mathbb{R}^{m}\rightarrow\mathbb{R}^{p}$ represent the nonlinear state dynamics and output equations, respectively. By discretizing the nonlinear state dynamics and accounting for the uncertainties in both state transition and measurements, \eqref{eq:ssa} can be represented in the following discrete-time stochastic form:
\begin{subequations}
\begin{align}
\bm x_k &= \bm \phi(\bm x_{k-1}, \bm u_{k-1}) + \bm w_k,
\qquad
\bm w_k \sim \mathcal{N}(\bm 0,\bm Q), \\
\bm z_k &= \bm h(\bm x_k, \bm u_k) + \bm v_k,
\qquad
\bm v_k \sim \mathcal{N}(\bm 0,\bm R), \\
\bm \mu_k &= \bm u_k + \bm \eta_k,
\qquad
\bm \eta_k \sim \mathcal{N}(\bm 0,\bm \Psi).
\end{align} 
\end{subequations}
where the function $\bm \phi:\mathbb{R}^{n}\times\mathbb{R}^{m}\rightarrow\mathbb{R}^{n}$ denotes the discrete state transition function. The actual system $\bm u_k $ is measured as $\bm \mu_k \!\in \! \mathbb{R}^m$ with uncertainty. $\bm w_k \!\in \!\mathbb{R}^n$, $\bm \eta_k \!\in \! \mathbb{R}^m$ and $\bm v_k \!\in \! \mathbb{R}^p$ denote the process noise and input/output measurement noise, respectively, and are assumed to be mutually independent zero-mean Gaussian random variables with covariance matrices $\bm Q$, $\bm R$ and $\bm \Psi$. The main goal of a decentralized DSE is to optimally estimate the internal states (and their corresponding covariances) at each timestep using port measurements in the presence of uncertainty.

\subsection{Stiffness-Aware Unscented Kalman Filtering} Starting from
the initial estimates $\bar{\bm x}_{0\vert 0}$ and
$\bm P_{0\vert 0}$, the filter recursively updates the state
estimates through prediction and correction steps.
\subsubsection{Prediction Step}
Given the mean $\bar{{\bm x}}_{k-1\vert k-1}$ and its corresponding covariance ${\bm P}_{k-1\vert k-1}$ corrected from the previous time step, the random variable $\bm s_{k-1\vert k-1}$ combining both state and input follows:
\begin{equation}
\begin{aligned}
\label{eq:rv}
{\bm s}_{k-1\vert k-1}&\sim
\mathcal{N}\!\left(
\begin{bmatrix}
\bar{\bm x}_{k-1\vert k-1} \\
\bm \mu_{k-1}
\end{bmatrix},
\begin{bmatrix}
{\bm P}_{k-1\vert k-1} & \bm 0 \\
\bm 0 & \bm \Psi
\end{bmatrix}
\right).
\end{aligned}
\end{equation}
Applying unscented transform \cite{Julier2000UT} to ${\bm s}_{k-1\vert k-1}$ yields $2(n+m)$ different sigma points $\bm{\mathsf s}^{(i)}_{k-1\vert k-1} \in \mathbb{R}^{n+m}$ with corresponding weights $w_i =1/2(n+m)$. Each sigma point is partitioned as,
\begin{align}
\label{eq:parti}
\bm{\mathsf s}^{(i)}_{k-1\vert k-1} =
\begin{bmatrix}
\bm {\mathsf x}^{(i)}_{k-1\vert k-1} \\
\bm {\mathsf u}^{(i)}_{k-1}
\end{bmatrix}, \quad i=1,\dots,2(n\!+\!m).
\end{align} 
where $\bm {\mathsf x}^{(i)}_{k-1\vert k-1} \in \mathbb{R}^{n}$ and $\bm {\mathsf u}^{(i)}_{k-1} \in \mathbb{R}^{m}$. Propagating the sigma points through $\bm f(\cdot)$ yields:
\begin{align}
\bm {\mathsf y}^{(i)}_{k} &= \bm f(\bm {\mathsf x}^{(i)}_{k-1\vert k-1}, \bm {\mathsf u}^{(i)}_{k-1}), \quad i=1,\dots,2(n\!+\!m).
\end{align}
Thus, we apply statistical linearization \cite{Sarkka2013} to $\bm f(\cdot)$ using the sigma points as follows,
\begin{subequations}
\allowdisplaybreaks
\begin{align}
\bm F_{k} &= \bm \Sigma_{\bm x \bm y,k} ^\top {\bm P}^{-1}_{k-1\vert k-1}, \label{sla}\\
\bm G_{k} &= \bm \Sigma_{\bm u \bm y,k} ^\top \bm \Psi^{-1},\\
\bm d_k &= \bar {\bm { y}}_k - \bm F_{k} \bar{\bm x}_{k-1\vert k-1} - \bm G_{k}\bar{\bm u}_{k-1}, \label{eq:sloffset}
\end{align}
\end{subequations}
where
\begin{subequations}
\allowdisplaybreaks
\begin{align}
\bar {\bm { y}}_k & = \sum_{i=1}^{2(n+m)} w_i
\bm{\mathsf y}^{(i)}_{k},\\
\bm \Sigma_{\bm x \bm y,k} & = \sum_{i=1}^{2(n+m)} w_i
\left(\bm {\mathsf x}^{(i)}_{k-1\vert k-1}\!-\!\bar{\bm x}_{k-1\vert k-1}\right)
\left(\bm{\mathsf y}^{(i)}_{k}\!-\!\bar{\bm y}_k\right)^\top, \\
\bm \Sigma_{\bm u \bm y,k} & = \sum_{i=1}^{2(n+m)}w_i
\left(\bm {\mathsf u}^{(i)}_{k}\!-\!\bar{\bm u}_{k}\right)
\left(\bm{\mathsf y}^{(i)}_{k}\!-\!\bar{\bm y}_k\right)^\top.
\end{align}
\end{subequations}
The linearization error $\bm e_k$ has zero mean due to \eqref{eq:sloffset}, and its covariance can be estimated as
\begin{equation}
\bm\Omega_k = \sum_{i=1}^{2(n+m)} w_i \bm{\mathsf e}^{(i)}_k \left(\bm{\mathsf e}^{(i)}_k\right)^\top
\end{equation}
where
\begin{equation}
\bm{\mathsf e}^{(i)}_k
=
\bm{\mathsf y}^{(i)}_{k}
-
\bm F_k \bm{\mathsf x}^{(i)}_{k}
-
\bm G_k \bm{\mathsf u}^{(i)}_{k}
-
\bm d_k.
\end{equation}
Therefore, statistical linearization yields the following local linear equivalent for the original nonlinear system:
\begin{equation}
\frac{\d \bm x(t)}{\d t}
=
\bm F_k \bm x(t)
+
\bm G_k \bm u(t)
+
\bm d_k
+
\bm e_k,
\quad \forall t \in (t_{k-1}, t_k].
\end{equation}
Assuming that $\bm u(t)=\bm u_{k-1}$ and $\bm e(t)=\bm e_k$ are constant over the interval $(t_{k-1}, t_k]$, define the augmented state as
\begin{equation}
\bm z(t)
=
\begin{bmatrix}
\bm x(t)\\
\bm u(t)\\
\bm d_k+\bm e_k
\end{bmatrix}.
\end{equation}
Then, the augmented system can be written as
\begin{equation}
\frac{\d \bm z(t)}{\d t}
=
\bm \Xi_k \bm z(t),
\end{equation}
with
\begin{equation}
\bm \Xi_k
=
\begin{bmatrix}
\bm F_k & \bm G_k & \bm I_{n\times n}\\
\bm 0_{m\times n} & \bm 0_{m\times m} & \bm 0_{m\times n}\\
\bm 0_{n\times n} & \bm 0_{n\times m} & \bm 0_{n\times n}
\end{bmatrix}.
\end{equation}
Over a sampling interval of length $h$, the exact discrete-time evolution is given by
\begin{equation}
\bm z_k
=
\exp(h\bm \Xi_k)\bm z_{k-1}.
\end{equation}
Partitioning $\exp(h\bm \Xi_k)$ yields
\begin{align}
\allowdisplaybreaks
{\setlength{\arraycolsep}{0pt}
\exp\left(h
\begin{bmatrix}
\bm F_k & \bm G_k & \bm I_{n\times n}\\
\bm 0_{m\times n} & \bm 0_{m\times m} & \bm 0_{m\times n}\\
\bm 0_{n\times n} & \bm 0_{n\times m} & \bm 0_{n\times n}
\end{bmatrix}\right)}\!=\!
{\setlength{\arraycolsep}{0pt}
\begin{bmatrix}
\bm \Phi_k & \bm \Gamma_k & \bm \Lambda_k\\
\bm 0_{m\times n} & \bm I_{m\times m} & \bm 0_{m\times n}\\
\bm 0_{n\times n} & \bm 0_{n\times m} & \bm I_{n\times n}
\end{bmatrix}}.
\label{eq:aug_exp}
\end{align}
Accordingly, the exact discrete-time equivalent of the original system is as follows
\begin{equation}
\bm x_k
=
\bm \Phi_k \bm x_{k-1}
+
\bm \Gamma_k \bm u_{k-1}
+
\bm \Lambda_k \bm d_k
+
\bm \Lambda_k \bm e_k.
\label{eq:sldiscrete}
\end{equation}
This yields the following time update equations for the state mean and covariance:
\begin{subequations}
\allowdisplaybreaks
\begin{align}
\bar{\bm x}_{k \vert k-1}
&=
\bm \Phi_k \bar{\bm x}_{k-1 \vert k-1}
+
\bm \Gamma_k \bm \mu_{k-1}
+
\bm \Lambda_k \bm d_k,
\label{eq:meanupdate}
\\
\bm P_{k \vert k-1}
&=
\bm \Phi_k \bm P_{k-1 \vert k-1} \bm \Phi_k^\top
+
\bm \Gamma_k \bm \Psi \bm \Gamma_k^\top
+
\bm \Lambda_k \bm \Omega_k \bm \Lambda_k^\top
+
\bm Q.
\label{eq:covupdate}
\end{align}
\end{subequations}

\subsubsection{Correction Step} The correction step is identical to that of the conventional UKF. Given $\bar {\bm x}_{k \vert k-1}$ and $\bm P_{k \vert k-1}$, $2(n+m)$ sigma points $\bm{\mathsf s}^{(i)}_{k\vert k-1} \in \mathbb{R}^{n+m}$ are generated from
\begin{equation}
\begin{aligned}
{\bm s}_{k\vert k-1}&\sim
\mathcal{N}\!\left(
\begin{bmatrix}
\bar{\bm x}_{k\vert k-1} \\
\bm \mu_{k}
\end{bmatrix},
\begin{bmatrix}
{\bm P}_{k\vert k-1} & \bm 0 \\
\bm 0 & \bm \Psi
\end{bmatrix}
\right),
\end{aligned}
\end{equation}
and are partitioned as,
\begin{align}
\bm{\mathsf s}^{(i)}_{k\vert k-1} =
\begin{bmatrix}
\bm {\mathsf x}^{(i)}_{k\vert k-1} \\
\bm {\mathsf u}^{(i)}_{k}
\end{bmatrix}, \quad i=1,\dots,2(n\!+\!m).
\end{align} 
where $\bm {\mathsf x}^{(i)}_{k\vert k-1} \in \mathbb{R}^{n}$ and $\bm {\mathsf u}^{(i)}_{k} \in \mathbb{R}^{m}$. Passing these sigma points through $\bm h(\cdot)$ yields:
\begin{align}
\bm {\mathsf z}^{(i)}_{{k\vert k-1}} &= \bm h(\bm {\mathsf x}^{(i)}_{k\vert k-1}, \bm {\mathsf u}^{(i)}_{k}), \quad i=1,\dots,2(n\!+\!m).
\end{align}
The predicted mean and covariance of output, and the cross variance between state and output are estimated as,
\begin{subequations}
\allowdisplaybreaks
\begin{align}
\bar {\bm { z}}_k & = \sum_{i=1}^{2(n+m)} w_i
\bm{\mathsf z}^{(i)}_{k},\\
\bm P_{\bm z \bm z,k} & = \sum_{i=1}^{2(n+m)} w_i
\left(\bm {\mathsf z}^{(i)}_{k\vert k-1}\!-\!\bar{\bm z}_{k\vert k-1}\right)
\left(\bm{\mathsf z}^{(i)}_{k\vert k-1}\!-\!\bar{\bm z}_{k\vert k-1}\right)^\top \notag\\ & \quad +\bm R,\\
 \bm P_{\bm x\bm z, k} &= \sum_{i=1}^{2(n+m)} w_i
\left(\bm{\mathsf x}^{(i)}_{k\vert k-1}\!-\!\bar{\bm x}_{k\vert k-1}\right) 
\left(\bm{\mathsf z}^{(i)}_{k\vert k-1}\!-\!\bar{\bm z}_{k\vert k-1}\right)^\top.
\end{align}
\end{subequations}
Finally, the state and covariance correction is expressed as,
\begin{subequations}
\allowdisplaybreaks
\begin{align}
\bm K_k &= \bm P_{\bm x\bm z,k} \bm P_{\bm z\bm z,k}^{-1},\\
\bar{\bm x}_{k\vert k} &= \bar{\bm x}_{k\vert k-1} +\bm K_k \left(\bm z_k - \bar{\bm z}_{k\vert k-1}\right),\\
{\bm P}_{k\vert k} &= {\bm P}_{k\vert k-1} -\bm K_k {\bm P}_{\bm z\bm z,k} \bm K_k^\top,
\end{align}
\end{subequations}
where $\bar{\bm x}_{k\vert k}$ and $ {\bm P}_{k\vert k} $ serve as the input of the next prediction step.\\ 
\noindent \textbf{Remark} 1) Statistical linearization should be understood as the best affine approximation in a minimum mean-square sense, rather than a Jacobian approximation; 
 2) for any continuous-time linear time-invariant dynamic system with piecewise-constant inputs, the matrix exponential discretization is exact, meanwhile the stability is preserved for arbitrary discretization step sizes; 
 3) the proposed method explicitly incorporates the covariance of the statistical linearization error into the covariance prediction \eqref{eq:covupdate}. This enables the filter to adapt its confidence in the process model based on the local linearization quality: when the linear approximation becomes less accurate, the increased error covariance shifts more weight to the measurement update, and vice versa. This formulation also distinguishes the contributions of intrinsic process noise, statistical linearization error, and input uncertainty in the covariance prediction, which is beneficial for parameter tuning.

\noindent\textbf{Theorem 1.} \textit{For continuous and differentiable $\bm f(\cdot)$, the one-step prediction errors of the state mean and covariance based on the statistically linearized model \eqref{eq:sldiscrete}, evaluated on the sigma-point-approximated distribution, are of second order in $h$, i.e., they both scale as $\mathcal O(h^2)$.}

\noindent\textit{Proof.} As the system input $\bm u_{k-1}$ is assumed to be constant over the sampling interval $[t_{k-1},t_k]$ of length $h$, the exact nonlinear system is given by
\begin{equation}
 \frac{\mathrm d \bm x(t)}{\mathrm d t}
 =
 \bm f(\bm x(t),\bm u_{k-1}).
 \label{eq:exact}
\end{equation}
The deterministic part of the linear surrogate model obtained from statistical linearization is
\begin{equation}
 \frac{\mathrm d \bm x(t)}{\mathrm d t}
 =
 \bm F_k\bm x(t)+\bm G_k\bm u_{k-1}+\bm d_k .
 \label{eq:linear_surrogate}
\end{equation}
For notational simplicity, define
\begin{subequations}
\allowdisplaybreaks
\begin{align}
 \bm x_k &\triangleq \bm x(t_k), \\ \bm x_{k-1} &\triangleq \bm x(t_{k-1}), \\
 \bm f_{k-1}
 &\triangleq
 \bm f(\bm x_{k-1},\bm u_{k-1}), \\
 \bm \ell_{k-1}
 &\triangleq
 \bm F_k\bm x_{k-1}+\bm G_k\bm u_{k-1}+\bm d_k .
\end{align}
\end{subequations}
Applying the Taylor expansion to \eqref{eq:exact} yields
\begin{equation}
 \bm x_k
 =
 \bm x_{k-1}
 +
 h \bm f_{k-1}
 +
 \mathcal O(h^2).
 \label{eq:x_exact_taylor}
\end{equation}
Similarly, applying the Taylor expansion to \eqref{eq:linear_surrogate} yields
\begin{equation}
 \Tilde{\bm x}_k
 =
 \bm x_{k-1}
 +
 h \bm \ell_{k-1}
 +
 \mathcal O(h^2).
 \label{eq:x_lin_taylor}
\end{equation}
Therefore, the one-step prediction error of the mean can be written as
\begin{equation}
\begin{aligned}
 \mathbb E[\Tilde{\bm x}_k]-\mathbb E[\bm x_k]
 &=
 h\Big(\mathbb E[\bm \ell_{k-1}] - \mathbb E[\bm f_{k-1}]\Big)
 +
 \mathcal O(h^2).
\end{aligned}
\end{equation}
Since statistical linearization enforces
\begin{equation}
 \mathbb E[\bm \ell_{k-1}] = \mathbb E[\bm f_{k-1}]
\end{equation}
under the adopted sigma-point approximation, it follows that
\begin{equation}
 \mathbb E[\Tilde{\bm x}_k]-\mathbb E[\bm x_k]
 =
 \mathcal O(h^2),
\end{equation}
which shows that the one-step prediction error of the mean is of second order in $h$.

To analyze the one-step prediction error of covariance, using \eqref{eq:x_exact_taylor}, the exact covariance can be expanded as
\begin{equation}
\begin{aligned}
\mathrm{Cov}(\bm x_k)
&=
\mathrm{Cov}\!\left(
\bm x_{k-1}+h\bm f_{k-1}+\mathcal O(h^2)
\right) \\
&=
\mathrm{Cov}(\bm x_{k-1})
+h\,\mathrm{Cov}(\bm x_{k-1},\bm f_{k-1}) \\
&\quad
+h\,\mathrm{Cov}(\bm f_{k-1},\bm x_{k-1})
+\mathcal O(h^2).
\end{aligned}
\label{eq:cov_exact_expand}
\end{equation}
Similarly, using \eqref{eq:x_lin_taylor}, the predicted covariance satisfies
\begin{equation}
\allowdisplaybreaks
\begin{aligned}
\mathrm{Cov}(\Tilde{\bm x}_k)
&=
\mathrm{Cov}\!\left(
\bm x_{k-1}+h\bm \ell_{k-1}+\mathcal O(h^2)
\right) \\
&=
\mathrm{Cov}(\bm x_{k-1})
+h\,\mathrm{Cov}(\bm x_{k-1},\bm \ell_{k-1}) \\
&\quad
+h\,\mathrm{Cov}(\bm \ell_{k-1},\bm x_{k-1})
+\mathcal O(h^2).
\end{aligned}
\label{eq:cov_lin_expand}
\end{equation}
Subtracting \eqref{eq:cov_exact_expand} from \eqref{eq:cov_lin_expand} gives
\begin{equation}
\allowdisplaybreaks
\begin{aligned}
\mathrm{Cov}(\Tilde{\bm x}_k)\!-\!\mathrm{Cov}(\bm x_k)
&=
h\,\mathrm{Cov}(\bm x_{k-1},\bm \ell_{k-1}) \\
&\quad
-h\,\mathrm{Cov}(\bm x_{k-1},\bm f_{k-1}) \\
&\quad
+h\,\mathrm{Cov}(\bm \ell_{k-1},\bm x_{k-1}) \\
&\quad
-h\,\mathrm{Cov}(\bm f_{k-1},\bm x_{k-1}) 
\!+\!\mathcal O(h^2).
\end{aligned}
\end{equation}
Since $\bm d_k$ is deterministic and $\bm u_{k-1}$ is assumed to be uncorrelated with $\bm x_{k-1}$, it follows that
\begin{equation}
\begin{aligned}
\mathrm{Cov}(\bm x_{k-1},\!\bm \ell_{k-1})
\!&=\!
\mathrm{Cov}(\bm x_{k-1},\!\bm F_k\bm x_{k-1}\!+\!\bm G_k\bm u_{k-1}\!+\!\bm d_k) \\
&=
\mathrm{Cov}(\bm x_{k-1},\bm F_k\bm x_{k-1}) \\
&=
\bm P_{k-1\vert k-1}\bm F_k^\top.
\end{aligned}
\end{equation}
Using \eqref{sla}, we obtain
\begin{equation}
\mathrm{Cov}(\bm x_{k-1},\bm \ell_{k-1})
=
\bm P_{k-1\vert k-1}\bm F_k^\top
=
\bm \Sigma_{\bm x\bm y,k}.
\end{equation}
By definition, $\bm \Sigma_{\bm x\bm y,k}$ is exactly the sigma-point approximation of
$\mathrm{Cov}(\bm x_{k-1},\bm f_{k-1})$. Hence,
\begin{equation}
\mathrm{Cov}(\bm x_{k-1},\bm \ell_{k-1})
=
\mathrm{Cov}(\bm x_{k-1},\bm f_{k-1})
\end{equation}
under the same sigma-point approximation. Likewise,
\begin{equation}
\mathrm{Cov}(\bm \ell_{k-1},\bm x_{k-1})
=
\mathrm{Cov}(\bm f_{k-1},\bm x_{k-1}).
\end{equation}
Therefore, the first-order covariance mismatch vanishes, and
\begin{equation}
\mathrm{Cov}(\Tilde{\bm x}_k)-\mathrm{Cov}(\bm x_k)
=
\mathcal O(h^2).
\end{equation}
Therefore, the one-step prediction error of covariance is also of second order in $h$.
\hfill $\blacksquare$ \\

\noindent\textbf{Remark.}
Errors introduced by the sigma-point moment matching generally do not vanish as $h$ decreases. As a result, such errors are not reflected in the $\mathcal O(h^2)$ scaling characterized in this theorem. This result implies that the proposed method preserves second-order prediction accuracy while improving numerical stability, providing a favorable trade-off between accuracy and robustness compared to explicit discretization methods.

\section{Numerical Results}

\label{numerical}
This section evaluates the proposed stiffness-aware unscented Kalman filter via numerical results. First, a toy single-machine infinite-bus (SMIB) system is used to illustrate and visualize the impact of numerical stability on conventional UKF based on explicit discretization. Then, a modified inverter-dominated IEEE 39-bus power system with both grid-following and grid-forming inverters is utilized to assess the effectiveness of the proposed method in a more realistic setting. 
\begin{figure}[!t]
 \centering
 \includegraphics[width=\linewidth]{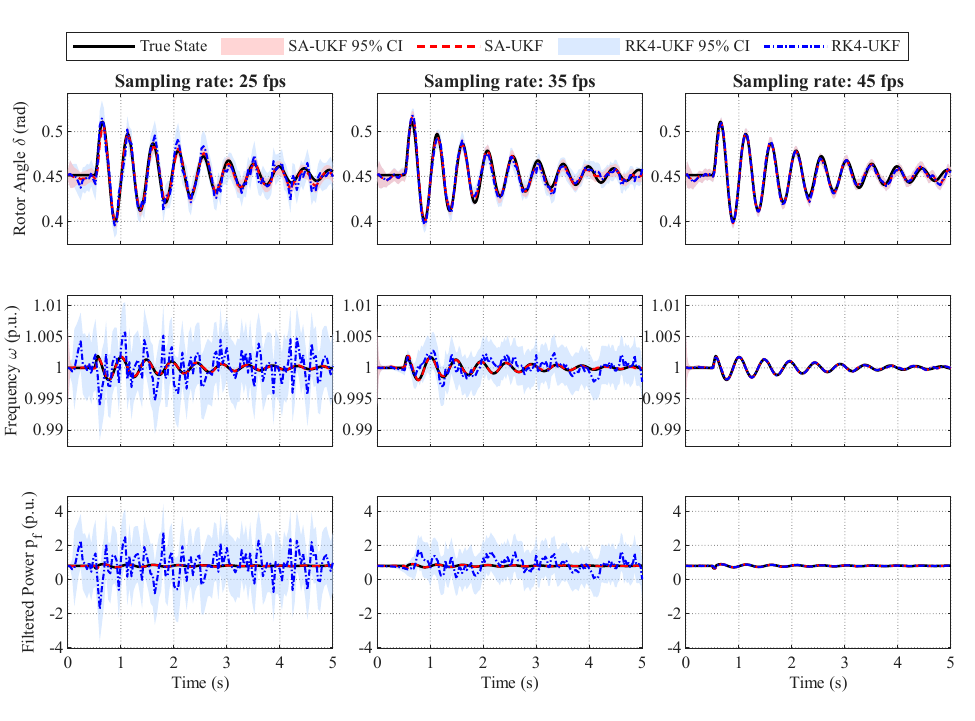}
 \caption{DSE trajectories of SA-UKF and RK4-UKF in the SMIB case under three sampling rates: the left, middle, and right columns correspond to 25, 35, and 45 fps, respectively; from top to bottom, the states are rotor angle $\delta$, frequency $\omega$, and filtered power $p_f$. The filtered state covariances are shown as 95\% confidence regions (CIs) using the shaded areas.}
 \label{fig:toy_results}
\end{figure}

\begin{figure}[!t]
 \centering
 \includegraphics[width=\linewidth]{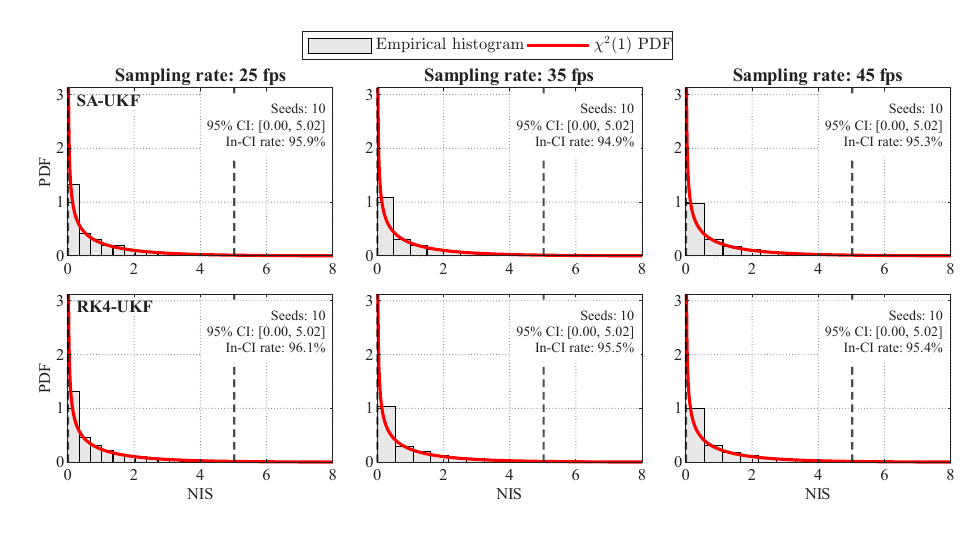}
 \caption{NIS consistency assessment (10 different random-seed selections) of the SA-UKF and RK4-UKF in the SMIB case under three sampling rates. The columns correspond to 25, 35, and 45 fps, and the two rows correspond to the SA-UKF and RK4-UKF, respectively.}
 \label{fig:toy_nis}
\end{figure}

\begin{figure}[!t]
 \centering
 \includegraphics[width=\columnwidth]{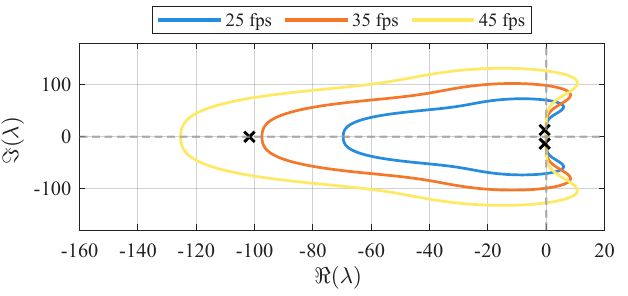}
 \caption{RK4 stability regions in the complex plane under three sampling rates.}
 \label{fig:toy_ros}
\end{figure}

\subsection{Toy Example: Single Machine Infinite Bus System}
In this single-machine infinite-bus (SMIB) system, the generation unit is modeled as a simplified virtual synchronous machine (VSM). 
\begin{subequations}
\allowdisplaybreaks
\begin{align}
 \frac{1}{\omega_b} \frac{\d \delta}{\d t} &= \omega-\omega_s \\
 H\frac{\d \omega}{\d t} &= p_\textnormal{ref} - p_f - D\left(\omega-\omega_s\right) \\
 \tau_p \frac{\d p_f}{\d t} &= -p_f + \frac{v_s v_g}{x}\sin \delta \label{eq:lpf}
\end{align}
\end{subequations}
\textit{Definition of Symbols}
\begin{itemize}
\item \textbf{State variables:}
\(\delta\) is the virtual rotor angle of the VSM with respect to the synchronous reference frame, \(\omega\) is the virtual frequency in per unit, and \(p_f\) is the filtered electromagnetic power.
\item \textbf{Parameters:}
\(\omega_b\) is the base angular frequency; \(\omega_s\) is the synchronous speed; \(H\) is the virtual inertia constant; \(D\) is the virtual damping coefficient; \(p_\textnormal{ref}\) is the active power reference; \(v_s\) is the internal virtual voltage magnitude; \(v_g\) is the infinite-bus voltage magnitude; \(x\) is the coupling reactance; and \(\tau_p\) is the time constant of the active-power low-pass filter.
\end{itemize}
The low-pass filter \eqref{eq:lpf} is introduced to capture the delay in the active-power measurement and processing path. Since its time scale is much faster than that of the swing dynamics, it introduces stiffness into the overall system.

\begin{table}[!t]
\centering
\caption{Default parameter setup for the SMIB case}
\label{tab:smib_params}
\renewcommand{\arraystretch}{1}
\setlength{\tabcolsep}{3pt}
\begin{tabular}{llp{5cm}}
\toprule
\textbf{Symbol} & \textbf{Value} & \textbf{Description} \\
\midrule
$\omega_b$ & $2\pi\times 60~\textnormal{rad/s}$ & Base angular frequency \\
$\omega_s$ & $1.0~\textnormal{p.u.}$ & Synchronous speed \\
$H$ & $3.5~\textnormal{s}$ & Virtual inertia constant \\
$D$ & $10.0$ & Virtual damping coefficient \\
$p_\textnormal{ref}$ & $0.8~\textnormal{p.u.}$ & Active power setpoint \\
$v_s$ & $1.1~\textnormal{p.u.}$ & Internal virtual voltage magnitude \\
$x$ & $0.6~\textnormal{p.u.}$ & Coupling reactance \\
$\tau_p$ & $0.01~\textnormal{s}$ & Active-power filter time constant \\
\bottomrule
\end{tabular}
\end{table}

For decentralized DSE, the infinite-bus voltage magnitude \(v_g\) is treated as the system input. The measurement output is taken as the true electrical active power
\begin{equation}
 p=\frac{v_s v_g}{x}\sin\delta.
\end{equation}
The default parameter setting of the SMIB system is given in Table~\ref{tab:smib_params}. In the simulation, the infinite-bus voltage magnitude with nominal value \( v_g =1.0\) p.u., is reduced to \(0.8\) p.u. at \(t=0.5\) s and then restored to its nominal value at \(t=0.55\) s to emulate a grid-side fault and its subsequent clearing. The measurement noise in both \(v_g\) and \(p\) is assumed to be zero-mean Gaussian with a standard deviation of \(1\times10^{-2}\) p.u. The proposed stiffness-aware UKF (SA-UKF) scheme is compared against the fourth-order Runge--Kutta-based UKF (RK4-UKF) scheme under three sampling rates, namely, \(25\), \(35\) and \(45\) fps. A detailed formulation of the fourth-order Runge--Kutta discretizations is provided in the Appendix. \ref{app:rk4}. 
The input/output measurement noise covariance matrices $\bm \Psi$ and $\bm R$ are chosen as $10^{-4}$, and
the process noise covariance matrix $\bm Q$ is chosen as $h^2\times10^{-6} \bm I$. The initial state mean is given by the steady state condition, while the initial state covariance is chosen as $1\times10^{-4} \bm I$.

\subsubsection{State Tracking Results}
The comparative results between these two DSE schemes are shown in Fig.\ref{fig:toy_results}. It is shown that the proposed SA-UKF maintains accurate and stable state tracking over all three sampling rates, whereas the RK4-UKF exhibits noticeably degraded estimation performance at lower sampling rates. At 45 fps, both methods closely match the true state. As the sampling rate decreases to 35 fps and especially 25 fps, the RK4-UKF produces larger oscillations and much wider confidence regions in the frequency and filtered-power estimates, whereas the SA-UKF remains well-behaved. These results verify the superior numerical robustness of the proposed SA-UKF under coarse sampling. In reality, the dominant slow dynamics of IBRs are often the primary concern for system operators. However, the standard UKF scheme may still require much higher sampling and computation rates to maintain numerical stability in the presence of multi-timescale dynamics, even though the slow dominant dynamics of interest can be adequately captured at a lower sampling rate.
\subsubsection{Statistical Consistency} The statistical consistency of the proposed SA-UKF and conventional RK4-UKF is verified by the normalized innovation square (NIS) represented as follows,
\begin{equation}
\operatorname{NIS}_k
=
\left(\bm z_k-\bar{\bm z}_{k\vert k-1}\right)^\top
\bm P_{\bm z\bm z,k}^{-1}
\left(\bm z_k-\bar{\bm z}_{k\vert k-1}\right).
\end{equation}
The filter is regarded as statistically consistent if the NIS approximately follows a chi-square distribution with \(p\) degrees of freedom, where \(p\) denotes the measurement dimension. For the SMIB case considered here, \(p=1\). The NIS consistency assessment across 10 random-seed selections is shown in Fig. \ref{fig:toy_nis}. The first row indicates that the proposed method is statistically consistent, further supporting the appropriateness of the chosen covariance matrices $\bm Q$, $\bm R$, and $\bm \Psi$. The RK4-UKF also appears statistically consistent under the NIS test. This is because NIS only evaluates the consistency between the innovation and its predicted covariance, but does not reflect the informativeness of the state estimate. When the sampling rate is insufficient, numerically induced oscillations may significantly inflate the state covariance, as shown in Fig.\ref{fig:toy_results}, so that the filter can still pass the NIS test even though the estimate has become overly uncertain and less informative. Therefore, such numerically induced oscillations may not be fully detected by the NIS test alone.

\subsubsection{Cause Analysis}
The underlying cause of the numerical oscillations observed in Fig.~\ref{fig:toy_results} can be explained by the RK4 stability function given in Appendix~\ref{app:rk4stability}. The corresponding regions of absolute stability under different sampling rates, together with the eigenvalues of the Jacobian matrix, are shown in Fig.~\ref{fig:toy_ros}. The SMIB system has three distinct eigenvalues: \(\lambda_{1,2} = -0.0055 \pm 0.1320 j\) and \(\lambda_3 = -101.77\). It can be seen that the two slow oscillatory modes \(\lambda_{1,2} \) remain inside the absolute stability region under all three sampling rates, while the fast real mode \(\lambda_3\) falls outside the stability region at 25 and 35 fps. This suggests that the observed numerical oscillations are mainly due to discretizing the fast mode rather than the slow, dominant dynamics of interest. For sampling rates below \(25\) fps, the RK4-UKF diverges and produces no meaningful results.

\subsection{Inverter-Dominated 39-Bus Power System}

\begin{figure}[!t]
 \centering
 \includegraphics[width=\columnwidth]{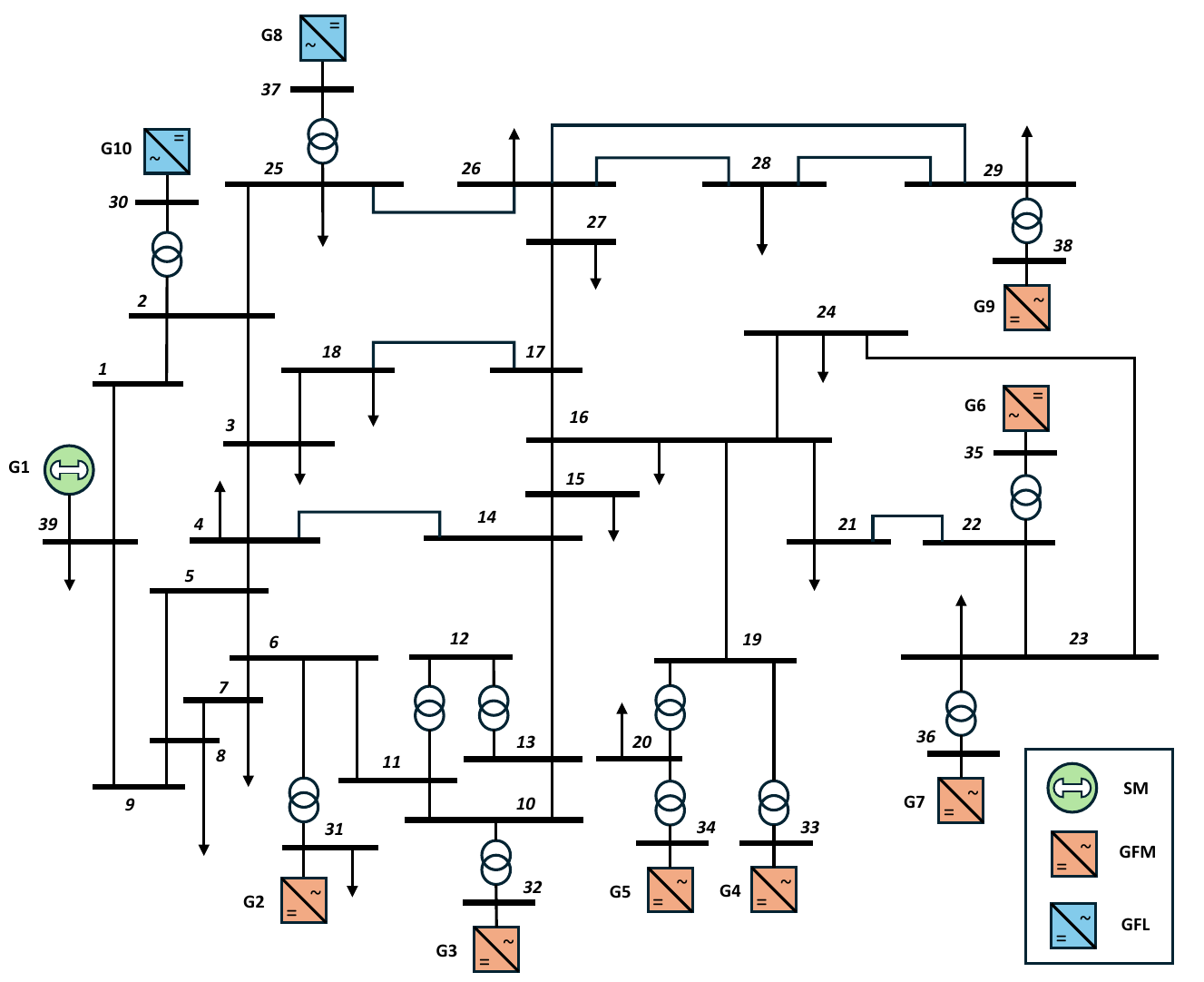}
 \caption{The modified inverter-dominated 39-bus power system }
 \label{fig:39bus}
\end{figure}

\begin{table}[!t]
\caption{Default parameters of the GFM and GFL models.}
\label{tab:ibr_params}
\centering
\begin{tabular}{lll}
\toprule
\textbf{Symbol} & \textbf{Value} & \textbf{Descriptions} \\
\midrule
\multicolumn{3}{l}{\textit{Common parameters}} \\
\midrule
$\omega_s$ & $1.0$ p.u. & System synchronous frequency \\
$\omega_b$ & $2\pi\times60$ & Base angular frequency \\
$r_f$ & $0.003$ p.u. & Converter-side filter resistance \\
$\ell_f$ & $0.08$ p.u. & Converter-side filter inductance \\
$c_f$ & $0.074$ p.u. & Filter capacitance \\
$r_g$ & $0.01$ p.u. & Grid-side resistance \\
$\ell_g$ & $0.2$ p.u. & Grid-side inductance \\
$k_p^c$ & $0.3771$ & Current-loop proportional gain \\
$k_i^c$ & $335.1032$ & Current-loop integral gain \\
\midrule
\multicolumn{3}{l}{\textit{GFM parameters}} \\
\midrule
$r_v$ & $0.0$ p.u. & Virtual resistance \\
$\ell_v$ & $0.2$ p.u. & Virtual inductance \\
$k_p$ & $0.1$ & Active-power droop gain \\
$k_q$ & $0.2$ & Reactive-power droop gain \\
$\omega_z$ & $1.0$ & Active-power measurement filter bandwidth \\
$\omega_f$ & $100.0$ & Reactive-power measurement filter bandwidth \\
$k_p^v$ & $0.3947$ & Voltage-loop proportional gain \\
$k_i^v$ & $49.5953$ & Voltage-loop integral gain \\
\midrule
\multicolumn{3}{l}{\textit{GFL parameters}} \\
\midrule
$k_p^{\mathrm{pll}}$ & $0.05$ & PLL proportional gain \\
$k_i^{\mathrm{pll}}$ & $1.42$ & PLL integral gain \\
$\omega_{lp}$ & $376.99$ & PLL low-pass filter cutoff frequency \\
$k_p^p$ & $0.05$ & Active-power loop proportional gain \\
$k_i^p$ & $0.6$ & Active-power loop integral gain \\
$k_p^q$ & $0.05$ & Reactive-power loop proportional gain \\
$k_i^q$ & $0.6$ p.u. & Reactive-power loop integral gain \\
$\omega_z$ & $41.47$ & Active-power measurement filter bandwidth \\
$\omega_f$ & $41.47$ & Reactive-power measurement filter bandwidth \\
\bottomrule
\end{tabular}
\end{table}
\begin{figure}[!t]
 \centering
 \includegraphics[width=\columnwidth]{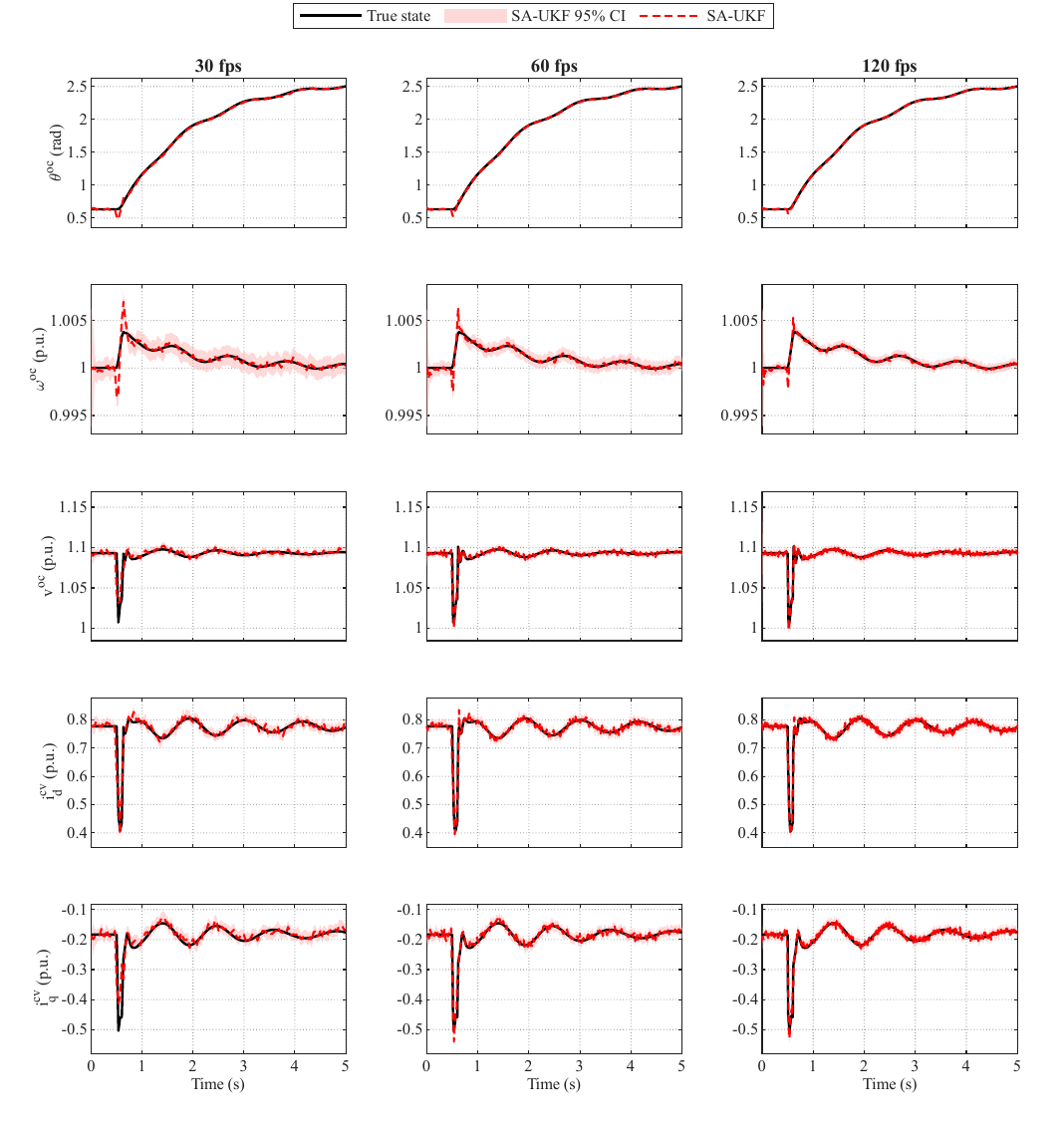}
 \caption{DSE trajectories of SA-UKF for $G8$ (a GFM) under three sampling rates: the left, middle and right columns correspond to 30, 60 and 120 fps, respectively; from top to bottom, the shown states are the virtual voltage phase angel $\theta^\textnormal{oc}$, the virtual frequency $\omega^\textnormal{oc}$, the virtual voltage magnitude $v^\textnormal{oc}$ and the converter-side $dq$-axis currents $i_d^\textnormal{cv}$ and $i_q^\textnormal{cv}$. The filtered state covariances are shown as 95\% confidence regions (CIs) using the shaded areas.}
 \label{fig:saukf_gfm}
\end{figure}

\begin{figure}[!t]
 \centering
 \includegraphics[width=\columnwidth]{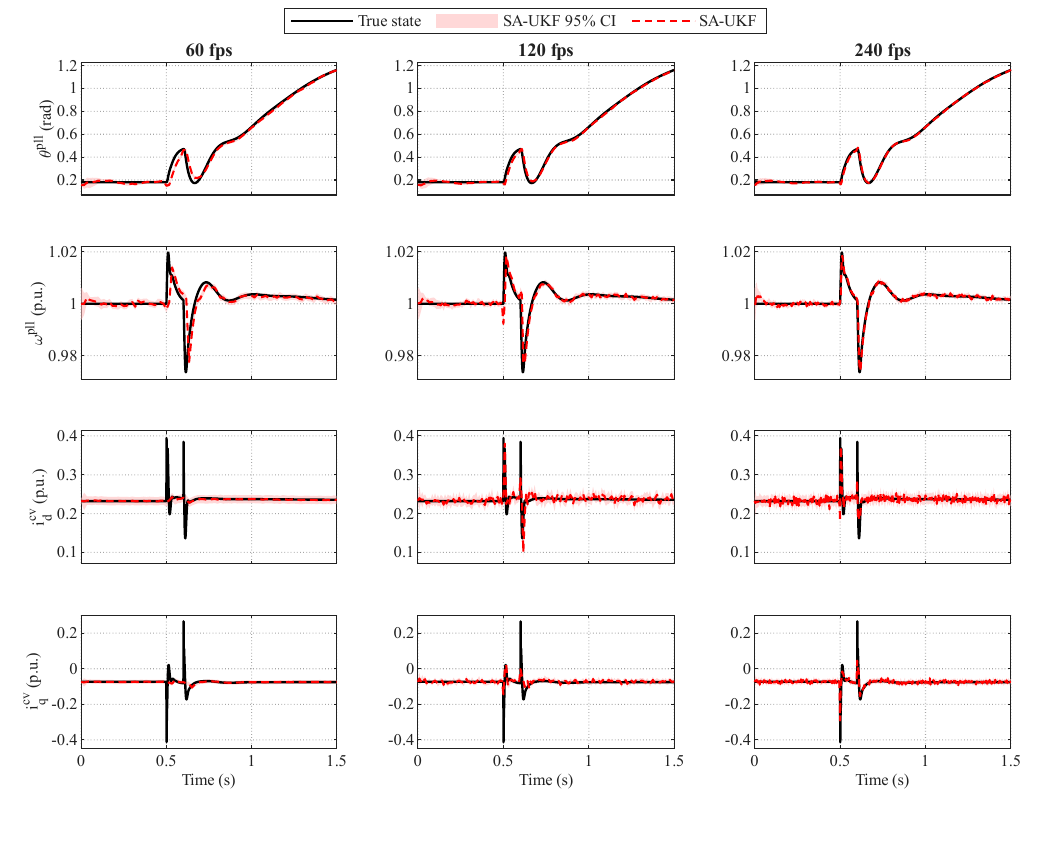}
 \caption{DSE trajectories of SA-UKF for $G10$ (a GFL) under three sampling rates: the left, middle and right columns correspond to 60, 120 and 240 fps, respectively; from top to bottom, the shown states are the PLL phase angel $\theta^\textnormal{pll}$, the PLL frequency $\omega^\textnormal{pll}$, and the converter-side $dq$-axis currents $i_d^\textnormal{cv}$ and $i_q^\textnormal{cv}$. The filtered state covariances are shown as 95\% confidence regions (CIs) using the shaded areas.}
 \label{fig:saukf_gfl}
\end{figure}

This modified 39-bus system has heterogeneous generation resources: $G1$ is a synchronous machine (SM) with a two-axis model, an IEEE DC1A exciter, and a TGOV1 turbine governor; $G8$ and $G10$ are grid-following inverters (GFLs); and the remaining generators are GFMs. The nonlinear state-space models of the GFLs and GFMs used in the decentralized DSE are provided in Appendix \ref{ibr_ssm}. The system inputs (voltage) are highlighted in blue, whereas the system outputs (current) are highlighted in red. Detailed derivations of the GFM and GFL input-output state-space models can be found in \cite{Xingyu:GFM} and \cite{Zhao:GFL:Note}, respectively. The default parameters of the GFMs and GFLs are given in \ref{tab:ibr_params}. A three-phase short-circuit fault at bus 21 is applied at $t=0.5$ s and later cleared at $t=0.6$ s. Assuming that the port measurements of $G9$ and $G10$ are available for decentralized DSE, we aim to estimate their internal dynamic states during the disturbance. The measurement noise in $v_{r}^{\textnormal{grid}}$, $v_{i}^{\textnormal{grid}}$, $i_{r}^{\textnormal{grid}}$ and $i_{i}^{\textnormal{grid}}$ are assumed to be zero-mean Gaussian with a standard deviation of \(1\times10^{-2}\) p.u. The input/output measurement noise covariance matrices $\bm \Psi$ and $\bm R$ are chosen as $10^{-4} \bm I$. The initial state mean is given by the steady state condition, while the initial state covariance is chosen as $1\times10^{-5} \bm I$. For GFMs, the process noise covariance matrix $\bm Q$ is set to $h^{2}\times10^{-4}\bm I$. For GFLs, $\bm Q$ is set to $h^{2}\times10^{-5}\bm I$ for the PLL and outer-loop states, and to $h^{2}\times10^{-1}\bm I$ for the inner-loop and filter states.

\subsubsection{Low-Rate State Tracking Results}

\begin{figure}[!t]
 \centering
 \includegraphics[width=\columnwidth]{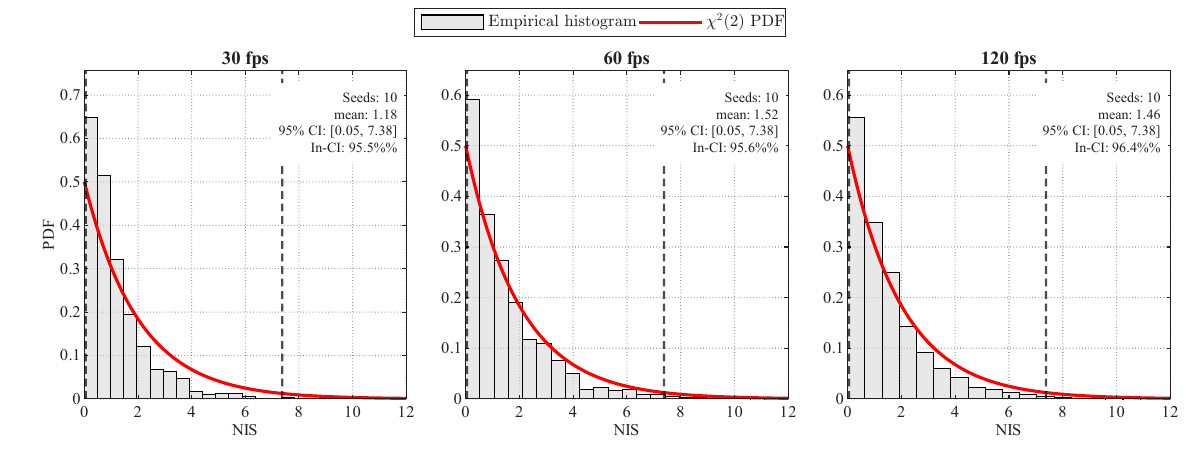}
 \caption{NIS consistency assessment (10 different random-seed selections) for SA-UKF on $G8$ (a GFM) under three sampling rates: 30, 60, and 120 fps respectively.}
 \label{fig:saukf_gfm_nis}
\end{figure}
\begin{figure}[!t]
 \centering
 \includegraphics[width=\columnwidth]{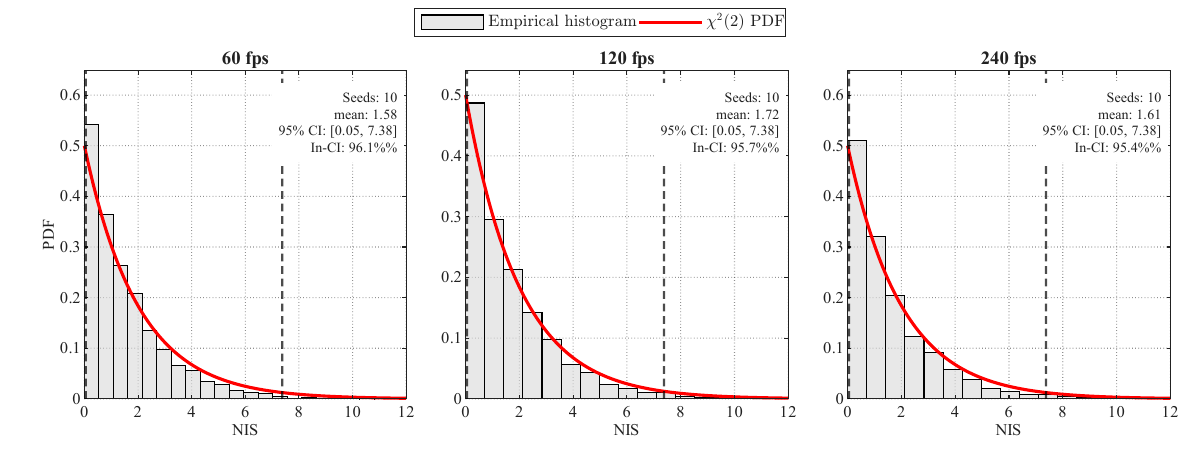}
 \caption{NIS consistency assessment (10 different random-seed selections) for SA-UKF on $G10$ (a GFL) under three different sampling rates: 60, 120, and 240 fps respectively.}
 \label{fig:saukf_gfl_nis}
\end{figure}
\begin{figure}[!t]
 \centering
 \includegraphics[width=\columnwidth]{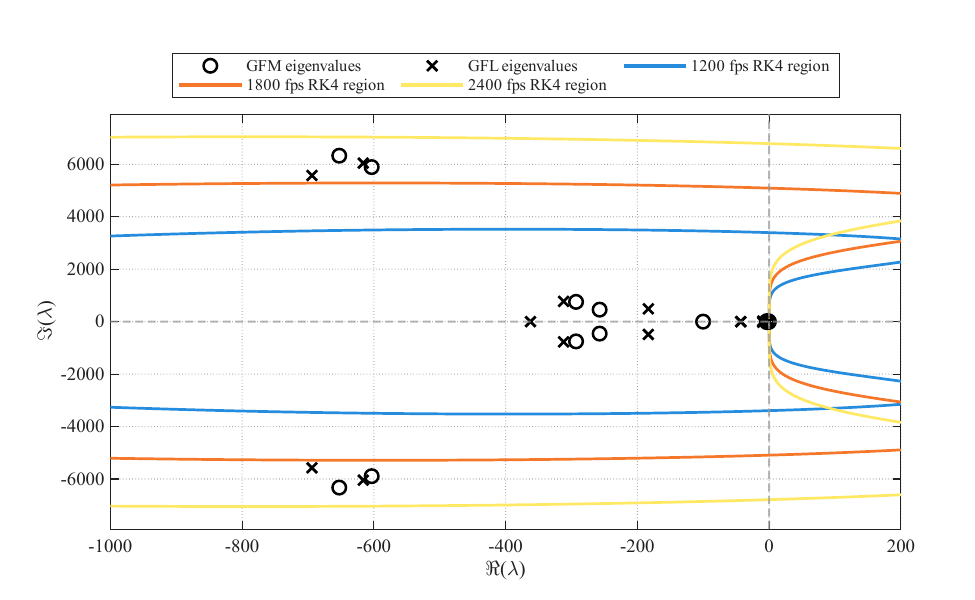}
 \caption{RK4 stability regions in the complex plane for different sampling rates, overlaid with the eigenvalues of the GFM and GFL dynamic equations.}
 \label{fig:rk4_ros}
\end{figure}
The numerical results of the proposed method for the decentralized DSE tasks of a GFM ($G8$) and a GFL ($G10$) are shown in Fig.~\ref{fig:saukf_gfm} and Fig.~\ref{fig:saukf_gfl}, respectively. The sampling rates range from $30$ fps to $240$ fps, consistent with standard synchrophasor reporting rates given in \cite{IEEE Std 60255}. In Fig.~\ref{fig:saukf_gfm}, the proposed method successfully tracks the GFM dynamic states under low sampling rates. The numerical results demonstrate that, when high-quality synchrophasor are available, using synchrophasor to estimate the dynamic states of a GFM is theoretically feasible. When the proposed method is applied to the GFL, the numerical results in Fig.~\ref{fig:saukf_gfl} show that the PLL-related states can be accurately tracked using synchrophasor, whereas the spikes in the converter-side current may not be fully captured due to the fundamental limitation of the sampling rate. Knowing that many oscillations caused by IBRs are associated with abnormal synchronization dynamics, the proposed method provides a feasible approach to monitor the synchronization process, such as the droop or PLL-related states, even under low sampling-rate measurements. The corresponding statistical consistency assessments are given in Fig.~\ref{fig:saukf_gfm_nis} and Fig.~\ref{fig:saukf_gfl_nis}. As the NISs both approximately follow a chi-square distribution with \(p=2\) degrees of freedom, the process noise covariance matrix $\bm Q$ is set appropriately. Since the proposed method can maintain statistical consistency under low sampling rates, its NIS can be used for anomaly detection, thereby providing addtional statistical information to support the real-time security of the power grid.

\begin{figure}[!t]
 \centering
 \includegraphics[width=\linewidth]{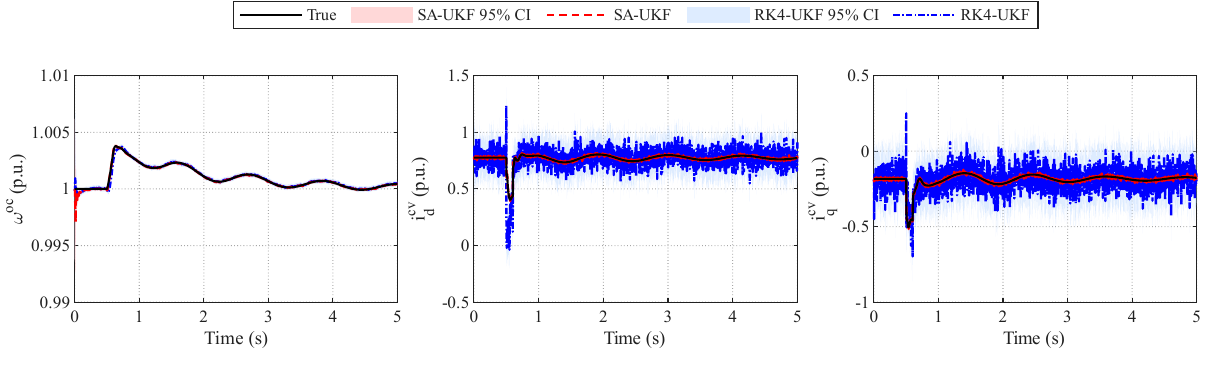}
 \caption{DSE results of SA-UKF and RK4-UKF for \(G8\) (a GFM) at 1800 fps sampling rate. The left, middle, and right columns show the virtual frequency \(\omega^{\mathrm{oc}}\) and the converter-side \(dq\)-axis currents \(i_d^{\mathrm{cv}}\) and \(i_q^{\mathrm{cv}}\), respectively. The filtered state covariances are shown as 95\% confidence intervals (CIs) using shaded regions.}
 \label{fig:boulder_gfm}
\end{figure}

\begin{figure}[!t]
 \centering
 \includegraphics[width=\linewidth]{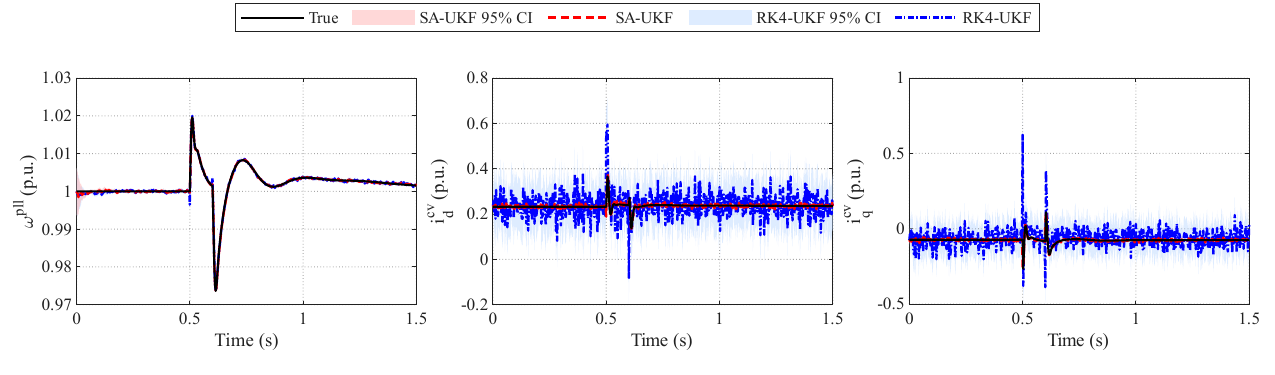}
 \caption{DSE results of SA-UKF and RK4-UKF for \(G10\) (a GFL) at 1800 fps sampling rate. The left, middle, and right columns show the PLL frequency \(\omega^{\mathrm{pll}}\) and the converter-side \(dq\)-axis currents \(i_d^{\mathrm{cv}}\) and \(i_q^{\mathrm{cv}}\), respectively. The filtered state covariances are shown as 95\% confidence intervals (CIs) using shaded regions.}
 \label{fig:boulder_gfl}
\end{figure}

\subsubsection{Stability Analysis of Explicit Discretization } The eigenvalues of the GFM and GFL process models along with the RK4 stability region are shown in Fig.~\ref{fig:rk4_ros}. At sampling rates of 1200 fps and 1800 fps, some eigenvalues remain outside the RK4 stability region, indicating potential numerical instability. Only when the sampling rate is increased to 2400 fps do all eigenvalues lie inside the RK4 stability region. This shows that conventional explicit-discretization-based methods require excessively high sampling rates to maintain numerical stability. As shown in Fig.~\ref{fig:boulder_gfm}, SA-UKF closely tracks the true trajectories of \(\omega^{\mathrm{oc}}\), \(i_d^{\mathrm{cv}}\), and \(i_q^{\mathrm{cv}}\). In contrast, RK4-UKF yields acceptable estimates for the slow state \(\omega^{\mathrm{oc}}\), but exhibits pronounced fluctuations and enlarged confidence regions in the fast current states, indicating the onset of numerical instability in the explicit prediction step. A similar trend is observed in Fig.~\ref{fig:boulder_gfl}. Although both methods capture the main transient of \(\omega^{\mathrm{pll}}\), RK4-UKF shows pronounced spurious oscillations in \(i_d^{\mathrm{cv}}\) and \(i_q^{\mathrm{cv}}\), whereas SA-UKF remains close to the true trajectories with tighter confidence regions. These results are consistent with Fig.~\ref{fig:rk4_ros}: when part of the eigenvalue spectrum lies outside the RK4 stability region, the explicit-discretization-based DSE becomes unreliable, especially for fast stiff states. By contrast, the proposed SA-UKF maintains both numerical robustness and estimation accuracy under the same sampling rate. When the sampling rate is below 1800 fps, the RK4 UKF diverges, and thus no meaningful estimation results can be obtained.

\subsubsection{Comparison with Implicit Discretization} Implicit methods are also commonly used for stiff dynamic models. Among them, the backward Euler method in Appendix \ref{app:BE} is particularly suitable for stiff model due to its \(L\)-stability. Therefore, as a benchmark, we implement a backward-Euler-based UKF, namely BE-UKF, in which the one-step prediction of each sigma point is obtained by solving the corresponding implicit equation using Newton iterations. With the Newton tolerance set to $10^{-10}$, the comparison between BE-UKF and SA-UKF is reported in Tables~\ref{tab:gfm_be_compare_single} and \ref{tab:gfl_be_compare_single}. The two methods achieve very similar estimation accuracy. However, the average computation time of BE-UKF is about 10 times higher than that of SA-UKF. Moreover, the computation time of BE-UKF may exceeds the sampling interval (highlighted in red),indicating real-time infeasibility. This is because, for each sigma point, BE-UKF must repeatedly evaluate the process model to approximate derivatives and to perform Newton iterations, leading to great amount of function evaluations. By contrast, SA-UKF requires only one function evaluation per sigma point at each step, and the propagates of mean and covariance are derived analytically. This makes the proposed method more efficient in computation and better suited for real-time applications.

\begin{table}[!t]
\centering
\caption{Comparison of RMSE and computational cost for the GFM.}
\label{tab:gfm_be_compare_single}
\setlength{\tabcolsep}{4.0pt}
\renewcommand{\arraystretch}{1.03}
\resizebox{\columnwidth}{!}{%
\begin{tabular}{cccccccc}
\toprule
\textbf{FPS} & \textbf{Mthd.} & $\omega^{\mathrm{oc}}$ (p.u.) & $i_d^{\mathrm{cv}}$ (p.u.) & \textbf{Avg. t (ms)} & \textbf{Max t (ms)} & \textbf{Avg. it./sp} & \textbf{Max it./sp} \\
\midrule
30 & SA-UKF & $5.02\times10^{-4}$ & $2.00\times10^{-2}$ & 0.306 & 1.80 & -- & -- \\
30 & BE-UKF & $4.77\times10^{-4}$ & $2.07\times10^{-2}$ & 4.38 & 18.5 & 3.86 & 4 \\
60 & SA-UKF & $3.16\times10^{-4}$ & $1.43\times10^{-2}$ & 0.257 & 1.55 & -- & -- \\
60 & BE-UKF & $2.86\times10^{-4}$ & $1.49\times10^{-2}$ & 3.63 & 8.76 & 3.36 & 4 \\
120 & SA-UKF & $2.10\times10^{-4}$ & $1.06\times10^{-2}$ & 0.248 & 1.54 & -- & -- \\
120 & BE-UKF & $1.86\times10^{-4}$ & $1.14\times10^{-2}$ & 2.80 & \textcolor{red}{11.6} & 3.00 & 4 \\
\bottomrule
\end{tabular}}
\end{table}

\begin{table}[!t]
\centering
\caption{Comparison of RMSE and computational cost for the GFL.}
\label{tab:gfl_be_compare_single}
\setlength{\tabcolsep}{4.0pt}
\renewcommand{\arraystretch}{1.03}
\resizebox{\columnwidth}{!}{%
\begin{tabular}{cccccccc}
\toprule
\textbf{FPS} & \textbf{Mthd.} & $\omega^{\mathrm{pll}}$ (p.u.) & $i_d^{\mathrm{cv}}$ (p.u.) & \textbf{Avg. t (ms)} & \textbf{Max t (ms)} & \textbf{Avg. it./sp} & \textbf{Max it./sp} \\
\midrule
60 & SA-UKF & $2.12\times10^{-3}$ & $3.71\times10^{-3}$ & 0.307 & 1.47 & -- & -- \\
60 & BE-UKF & $2.03\times10^{-3}$ & $5.01\times10^{-3}$ & 4.03 & 10.0 & 3.38 & 4 \\
120 & SA-UKF & $1.45\times10^{-3}$ & $8.29\times10^{-3}$ & 0.305 & 1.92 & -- & -- \\
120 & BE-UKF & $1.33\times10^{-3}$ & $7.25\times10^{-3}$ & 3.54 & \textcolor{red}{21.4} & 3.10 & 4 \\
240 & SA-UKF & $6.49\times10^{-4}$ & $7.07\times10^{-3}$ & 0.259 & 1.91 & -- & -- \\
240 & BE-UKF & $7.37\times10^{-4}$ & $6.51\times10^{-3}$ & 2.94 & \textcolor{red}{11.0} & 3.00 & 4 \\
\bottomrule
\end{tabular}}
\end{table}

\section{Conclusion}
This paper presents a derivative-free and stiffness-aware decentralized DSE method for inverter-dominated power systems with stiff dynamics under insufficient sampling rates. By combining statistical linearization with matrix-exponential discretization, the proposed method replaces the conventional explicit prediction step with a Jacobian-free scheme that maintains numerical robustness under coarse sampling. Theoretical analysis shows that the one-step prediction errors of the state mean and covariance are both second order in the sampling interval. Numerical studies on the SMIB system and the inverter-dominated IEEE 39-bus system demonstrate that the proposed method can provide stable and accurate state estimation at sampling rates where explicit-discretization-based DSE becomes unreliable, while also requiring much lower computational cost than implicit-discretization-based alternatives. %\sout{Therefore, the proposed method offers a practical solution for the situational awareness of IBR-dominated power systems using PMU data or heavily down-sampled waveform measurements.} 
The proposed stiffness-aware DSE framework provides a practical pathway for monitoring inverter-dominated power systems using PMU data, enabling improved situational awareness, anomaly detection, and stability assessment under realistic measurement constraints. 

Future work will extend this framework to joint state and parameter estimation, anomaly detection, and online monitoring of high-frequency oscillations.

\appendices
\section{The Fourth-Order Runge--Kutta Scheme}
\subsection{Formulation}
\label{app:rk4}
Given the continuous-time process model in \eqref{eq:exact},
the fourth-order Runge--Kutta (RK4) discretization is given by
\begin{subequations}
\allowdisplaybreaks
\begin{align}
\bm k_1 &= \bm f(\bm x_{k-1}, \bm u_{k-1}),\\
\bm k_2 &= \bm f\!\left(\bm x_{k-1}+\frac{h}{2}\bm k_1, \bm u_{k-1}\right),\\
\bm k_3 &= \bm f\!\left(\bm x_{k-1}+\frac{h}{2}\bm k_2, \bm u_{k-1}\right),\\
\bm k_4 &= \bm f\!\left(\bm x_{k-1}+h\bm k_3, \bm u_{k-1}\right),\\
\bm x_k &= \bm x_{k-1}+\frac{h}{6}\left(\bm k_1+2\bm k_2+2\bm k_3+\bm k_4\right).
\end{align}
\end{subequations}
This discrete-time update defines the state-transition mapping $\bm \phi(\cdot)$ used in the standard UKF, leading to the RK4-UKF.

\subsection{Region of Absolute Stability}
\label{app:rk4stability}
For a nonlinear system, the local numerical stability of the RK4 discretization can be assessed from the eigenvalues $\lambda_i$ of the Jacobian
\begin{equation}
 \mathbf J
 =
 \left.
 \frac{\partial \bm f(\bm x,\bm u)}{\partial \bm x}
 \right|_{(\bm x^\star,\bm u^\star)}.
\end{equation}
RK4 is numerically stable only if all scaled eigenvalues $h\lambda_i$ lie inside its absolute-stability region. This region is characterized by the stability function
\begin{equation}
 R(z)=1+z+\frac{z^2}{2}+\frac{z^3}{6}+\frac{z^4}{24},
\end{equation}
with $z=h\lambda$, and is defined as
\begin{equation}
 \mathcal S_{\mathrm{RK4}}
 =
 \left\{
 z\in\mathbb{C}: |R(z)|\le 1
 \right\}.
\end{equation}
Accordingly, the RK4 discretization is locally stable only if
\begin{equation}
 h\lambda_i \in \mathcal S_{\mathrm{RK4}}, \qquad \forall i.
\end{equation}

\section{The Backward-Euler Scheme}
\label{app:BE}
The backward Euler discretization is expressed as
\begin{equation}
 \bm x_k = \bm x_{k-1} + h\, \bm f(\bm x_k, \bm u_{k-1}),
\end{equation}
where the implicit state $\bm x_k$ is solved iteratively via Newton's method:
\begin{subequations}
\begin{align}
 \bm r_k^{(j)}
 &= \bm x_k^{(j)}-\bm x_{k-1}
 - h\,\bm f\!\left(\bm x_k^{(j)},\bm u_{k-1}\right), \\
 \bm x_k^{(j+1)}
 &= \bm x_k^{(j)}
 - \left[
 \bm I-h\,\frac{\partial \bm f}{\partial \bm x}
 \Big(\bm x_k^{(j)},\bm u_{k-1}\Big)
 \right]^{-1}
 \bm r_k^{(j)}.
\end{align}
\end{subequations}
In each iteration, the Jacobian matrix is approximated by numerical forward differentiation. The converged implicit iteration defines the state-transition mapping $\bm \phi(\cdot)$ used in the BE-UKF.

\section{Input-Output State-Space Models of IBRs}
\label{ibr_ssm}
\subsection*{Grid-Forming Inverter}
\subsubsection*{Process Model}
\begin{subequations}\label{gfm_ode}
\allowdisplaybreaks
\begin{align}
\frac{1}{\omega_{b}} \frac{\d \theta^\textnormal{oc}}{\d t}
&= \omega^\textnormal{oc} \!-\! \omega_s
\\[0.5ex]
\frac{1}{\omega_{z}k_p}\frac{\d \omega^\textnormal{oc}}{\d t}
&= p^\textnormal{ref} \!-\! v_{r}^{\textnormal{filt}}i_{r}^{\textnormal{filt}} \!-\! v_{i}^{\textnormal{filt}}i_{i}^{\textnormal{filt}}
\!+\! \frac{1}{k_p}\bigl(\omega^\textnormal{ref} \!-\! \omega^\textnormal{oc}\bigr)
\\[0.5ex]
\frac{1}{\omega_{f}k_q}\frac{\d v^\textnormal{oc}}{\d t}
&= q^\textnormal{ref} \!-\! v_{i}^{\textnormal{filt}}i_{r}^{\textnormal{filt}} \!+\! v_{r}^{\textnormal{filt}}i_{i}^{\textnormal{filt}}
\!+\! \frac{1}{k_q}(v^\textnormal{ref} \!-\! v^\textnormal{oc})
\\[0.5ex]
\frac{\ell_{f}}{\omega_{b}} \frac{\d i_{d}^{\textnormal{cv}}}{\d t}
&=
k_{p}^{c} \Bigl(
k_{p}^{v} \bigl(
v^\textnormal{oc}
\!-\! r_v i_{d}^{\textnormal{filt}}
\!+\! \omega^\textnormal{oc} \ell_v i_{q}^{\textnormal{filt}}
\!-\! v_{d}^{\textnormal{filt}}
\bigr)
\nonumber\\
&\qquad
\!+\! k_{i}^{v}\xi_{d}
\!-\! c_f \omega^{\textnormal{oc}} v_{q}^{\textnormal{filt}}
\!-\! i_{d}^{\textnormal{cv}}
\Bigr)
\nonumber\\
&\qquad
\!+\! k_{i}^{c}\gamma_{d}
\!-\! v_{d}^{\textnormal{filt}}
\!-\! r_{f} i_{d}^{\textnormal{cv}}
\\[0.5ex]
\frac{\ell_{f}}{\omega_{b}} \frac{\d i_{q}^{\textnormal{cv}}}{\d t}
&=
k_{p}^{c} \Bigl(
k_{p}^{v} \bigl(
\!-\! r_v i_{q}^{\textnormal{filt}}
\!-\! \omega^\textnormal{oc} \ell_v i_{d}^{\textnormal{filt}}
\!-\! v_{q}^{\textnormal{filt}}
\bigr)
\nonumber\\
&\qquad
\!+\! k_{i}^{v}\xi_{q}
\!+\! c_f \omega^{\textnormal{oc}} v_{d}^{\textnormal{filt}}
\!-\! i_{q}^{\textnormal{cv}}
\Bigr)
\nonumber\\
&\qquad
\!+\! k_{i}^{c}\gamma_{q}
\!-\! v_{q}^{\textnormal{filt}}
\!-\! r_{f} i_{q}^{\textnormal{cv}}
\\[0.5ex]
\frac{c_{f}}{\omega_{b}} \frac{\d v_{d}^{\textnormal{filt}}}{\d t}
&=
i_{d}^{\textnormal{cv}}
\!-\! i_{d}^{\textnormal{filt}}
\!+\! \omega^{\textnormal{oc}} c_{f} v_{q}^{\textnormal{filt}}
\\[0.5ex]
\frac{c_{f}}{\omega_{b}} \frac{\d v_{q}^{\textnormal{filt}}}{\d t}
&=
i_{q}^{\textnormal{cv}}
\!-\! i_{q}^{\textnormal{filt}}
\!-\! \omega^{\textnormal{oc}} c_{f} v_{d}^{\textnormal{filt}}
\\[0.8ex]
\frac{\ell_{g}}{\omega_{b}} \frac{\d i_{d}^{\textnormal{filt}}}{\d t}
&=
v_{d}^{\textnormal{filt}}
\!-\!
\sin{\left(\theta^{\textnormal{oc}}\!+\!\frac{\pi}{2}\right)}
{\color{blue}{v_{r}^{\textnormal{grid}}}}
\!+\!
\cos{\left(\theta^{\textnormal{oc}}\!+\!\frac{\pi}{2}\right)}
{\color{blue}{v_{i}^{\textnormal{grid}}}}
\nonumber\\
&\qquad
\!-\! r_{g} i_{d}^{\textnormal{filt}}
\!+\! \omega^{\textnormal{oc}} \ell_{g} i_{q}^{\textnormal{filt}}
\\[1.0ex]
\frac{\ell_{g}}{\omega_{b}} \frac{\d i_{q}^{\textnormal{filt}}}{\d t}
&=
v_{q}^{\textnormal{filt}}
\!-\!
\cos{\left(\theta^{\textnormal{oc}}\!+\!\frac{\pi}{2}\right)}
{\color{blue}{v_{r}^{\textnormal{grid}}}}
\!-\!
\sin{\left(\theta^{\textnormal{oc}}\!+\!\frac{\pi}{2}\right)}
{\color{blue}{v_{i}^{\textnormal{grid}}}}
\nonumber\\
&\qquad
\!-\! r_{g} i_{q}^{\textnormal{filt}}
\!-\! \omega^{\textnormal{oc}} \ell_{g} i_{d}^{\textnormal{filt}}
\\[0.5ex]
\frac{\d \xi_{d}}{\d t}
&=
v^\textnormal{oc}
\!-\! r_v i_{d}^{\textnormal{filt}}
\!+\! \omega^\textnormal{oc} \ell_v i_{q}^{\textnormal{filt}}
\!-\! v_{d}^{\textnormal{filt}}
\\[0.5ex]
\frac{\d \xi_{q}}{\d t}
&=
\!-\! r_v i_{q}^{\textnormal{filt}}
\!-\! \omega^\textnormal{oc} \ell_v i_{d}^{\textnormal{filt}}
\!-\! v_{q}^{\textnormal{filt}}
\\[0.5ex]
\frac{\d \gamma_{d}}{\d t}
&=
k_{p}^{v} \bigl(
v^\textnormal{oc}
\!-\! r_v i_{d}^{\textnormal{filt}}
\!+\! \omega^\textnormal{oc} \ell_v i_{q}^{\textnormal{filt}}
\!-\! v_{d}^{\textnormal{filt}}
\bigr)
\nonumber\\
&\qquad
\!+\! k_{i}^{v}\xi_{d}
\!-\! c_f \omega^{\textnormal{oc}} v_{q}^{\textnormal{filt}}
\!-\! i_{d}^{\textnormal{cv}}
\\[0.5ex]
\frac{\d \gamma_{q}}{\d t}
&=
k_{p}^{v} \bigl(
\!-\! r_v i_{q}^{\textnormal{filt}}
\!-\! \omega^\textnormal{oc} \ell_v i_{d}^{\textnormal{filt}}
\!-\! v_{q}^{\textnormal{filt}}
\bigr)
\nonumber\\
&\qquad
\!+\! k_{i}^{v}\xi_{q}
\!+\! c_f \omega^{\textnormal{oc}} v_{d}^{\textnormal{filt}}
\!-\! i_{q}^{\textnormal{cv}}
\end{align}
\end{subequations}
\subsubsection*{Measurement Model}
\begin{subequations}\label{gfm_meas}
\begin{align}
{\color{red} i^\textnormal{filt}_{r}}
&=
\sin\!\left(\theta^\textnormal{oc} \!+\! \frac{\pi}{2}\right) i^\textnormal{filt}_{d}
\!+\!
\cos\!\left(\theta^\textnormal{oc} \!+\! \frac{\pi}{2}\right) i^\textnormal{filt}_{q}
\\[0.5ex]
{\color{red} i^\textnormal{filt}_{i}}
&=
-\cos\!\left(\theta^\textnormal{oc} \!+\! \frac{\pi}{2}\right) i^\textnormal{filt}_{d}
\!+\!
\sin\!\left(\theta^\textnormal{oc} \!+\! \frac{\pi}{2}\right) i^\textnormal{filt}_{q}
\end{align}
\end{subequations}

\subsection*{Grid-Following Inverter}
\subsubsection*{Process Model}

\begin{subequations}\label{gfl_ode}
\allowdisplaybreaks
\begin{align}
\frac{1}{\omega_{b}} \frac{\d \theta^\textnormal{pll}}{\d t}
&= \omega^\textnormal{pll} \!-\! \omega_s
\\[0.5ex]
\frac{1}{\omega_{lp}} \frac{\d \omega^\textnormal{pll}}{\d t}
&=
\left(
\frac{k_i^\textnormal{pll}}{\omega_{lp}} \!-\! k_p^\textnormal{pll}
\right) v_q^\textnormal{pll}
\!+\!
k_p^\textnormal{pll} v_q^\textnormal{filt}
\\[0.5ex]
\frac{1}{\omega_{lp}} \frac{\d v_q^\textnormal{pll}}{\d t}
&=
- v_q^\textnormal{pll}
\!+\!
v_q^\textnormal{filt}
\\[0.5ex]
\frac{\d \sigma_p}{\d t}
&=
p^\textnormal{ref}
\!-\!
p_m
\\[0.5ex]
\frac{\d \sigma_q}{\d t}
&=
q^\textnormal{ref}
\!-\!
q_m
\\[0.5ex]
\frac{1}{\omega_z} \frac{\d p_m}{\d t}
&=
v_d^\textnormal{filt} i_d^\textnormal{filt}
\!+\!
v_q^\textnormal{filt} i_q^\textnormal{filt}
\!-\!
p_m
\\[0.5ex]
\frac{1}{\omega_f} \frac{\d q_m}{\d t}
&=
- v_d^\textnormal{filt} i_q^\textnormal{filt}
\!+\!
v_q^\textnormal{filt} i_d^\textnormal{filt}
\!-\!
q_m
\\[0.5ex]
\frac{\d \gamma_d}{\d t}
&=
k_p^p \bigl(p^\textnormal{ref} \!-\! p_m\bigr)
\!+\!
k_i^p \sigma_p
\!-\!
i_d^\textnormal{cv}
\\[0.5ex]
\frac{\d \gamma_q}{\d t}
&=
-
k_p^q \bigl(q^\textnormal{ref} \!-\! q_m\bigr)
\!-\!
k_i^q \sigma_q
\!-\!
i_q^\textnormal{cv}
\\[0.5ex]
\frac{\ell_f}{\omega_b} \frac{\d i_d^\textnormal{cv}}{\d t}
&=
k_p^c \Bigl(
k_p^p \bigl(p^\textnormal{ref} \!-\! p_m\bigr)
\!+\!
k_i^p \sigma_p
\!-\!
i_d^\textnormal{cv}
\Bigr)
\nonumber\\
&\qquad
\!+\!
k_i^c \gamma_d
\!-\!
v_d^\textnormal{filt}
\!-\!
r_f i_d^\textnormal{cv}
\\[0.5ex]
\frac{\ell_f}{\omega_b} \frac{\d i_q^\textnormal{cv}}{\d t}
&=
k_p^c \Bigl(
-
k_p^q \bigl(q^\textnormal{ref} \!-\! q_m\bigr)
\!-\!
k_i^q \sigma_q
\!-\!
i_q^\textnormal{cv}
\Bigr)
\nonumber\\
&\qquad
\!+\!
k_i^c \gamma_q
\!-\!
v_q^\textnormal{filt}
\!-\!
r_f i_q^\textnormal{cv}
\\[0.5ex]
\frac{c_f}{\omega_b} \frac{\d v_d^\textnormal{filt}}{\d t}
&=
i_d^\textnormal{cv}
\!-\!
i_d^\textnormal{filt}
\!+\!
\omega^\textnormal{pll} c_f v_q^\textnormal{filt}
\\[0.5ex]
\frac{c_f}{\omega_b} \frac{\d v_q^\textnormal{filt}}{\d t}
&=
i_q^\textnormal{cv}
\!-\!
i_q^\textnormal{filt}
\!-\!
\omega^\textnormal{pll} c_f v_d^\textnormal{filt}
\\[0.8ex]
\frac{\ell_g}{\omega_b} \frac{\d i_d^\textnormal{filt}}{\d t}
&=
v_d^\textnormal{filt}
\!-\!
\sin\!\left(\theta^\textnormal{pll}\!+\!\frac{\pi}{2}\right)
{\color{blue}{v_r^\textnormal{grid}}}
\!+\!
\cos\!\left(\theta^\textnormal{pll}\!+\!\frac{\pi}{2}\right)
{\color{blue}{v_i^\textnormal{grid}}}
\nonumber\\
&\qquad
\!-\!
r_g i_d^\textnormal{filt}
\!+\!
\omega^\textnormal{pll} \ell_g i_q^\textnormal{filt}
\\[1.0ex]
\frac{\ell_g}{\omega_b} \frac{\d i_q^\textnormal{filt}}{\d t}
&=
v_q^\textnormal{filt}
\!-\!
\cos\!\left(\theta^\textnormal{pll}\!+\!\frac{\pi}{2}\right)
{\color{blue}{v_r^\textnormal{grid}}}
\!-\!
\sin\!\left(\theta^\textnormal{pll}\!+\!\frac{\pi}{2}\right)
{\color{blue}{v_i^\textnormal{grid}}}
\nonumber\\
&\qquad
\!-\!
r_g i_q^\textnormal{filt}
\!-\!
\omega^\textnormal{pll} \ell_g i_d^\textnormal{filt}
\end{align}
\end{subequations}

\subsubsection*{Measurement Model}

\begin{subequations}\label{gfl_meas}
\begin{align}
{\color{red} i^\textnormal{filt}_{r}}
&=
\sin\!\left(\theta^\textnormal{pll} \!+\! \frac{\pi}{2}\right) i^\textnormal{filt}_{d}
\!+\!
\cos\!\left(\theta^\textnormal{pll} \!+\! \frac{\pi}{2}\right) i^\textnormal{filt}_{q}
\\[0.5ex]
{\color{red} i^\textnormal{filt}_{i}}
&=
-\cos\!\left(\theta^\textnormal{pll} \!+\! \frac{\pi}{2}\right) i^\textnormal{filt}_{d}
\!+\!
\sin\!\left(\theta^\textnormal{pll} \!+\! \frac{\pi}{2}\right) i^\textnormal{filt}_{q}
\end{align}
\end{subequations}


\begin{thebibliography}{99}
\bibitem{NERC:PLL}“1,200 MW fault induced solar photovoltaic resource interruption disturbance
report: Southern California 8/16/2016 event,” North American Electric Reliability Corporation (NERC), Tech. Rep., Jun. 2017.
\bibitem{fan2022currentlimit}
B.~Fan, T.~Liu, F.~Zhao, H.~Wu, and X.~Wang, ``A review of current-limiting control of grid-forming inverters under symmetrical disturbances,'' \emph{IEEE Open J. Power Electron.}, vol.~3, pp.~955--969, 2022.
\bibitem{Savastianov:MC}M. Savastianov, K. Smedley and J. Cao, “Power system recovery from momentary cessation with transient stability improvement," {\em IEEE Trans. Power Syst.}, 2023.
\bibitem{marchi2023dse}
P.~Marchi, P.~Gill~Estevez, and C.~Galarza, ``DSE-assisted DEF strategy for locating forced oscillations in synchronous generators,'' \emph{Int. J. Electr. Power Energy Syst.}, vol.~146, p.~108744, Mar.~2023, doi: 10.1016/j.ijepes.2022.108744.
\bibitem{Bamigbade:cyber} A. Bamigbade, Y. Dvorkin and R. Karri, ``Cyberattack on phase-locked loops in inverter-based energy resources," {\em IEEE Trans. Smart Grid}, vol. 15, no. 1, pp. 821-833, Jan. 2024.
\bibitem{zhao2018robustdecentralized}
J.~B.~Zhao and L.~Mili, ``Power system robust decentralized dynamic state estimation based on multiple hypothesis testing,'' \emph{IEEE Trans. Power Syst.}, vol.~33, no.~4, pp.~4553--4562, Jul.~2018, doi: 10.1109/TPWRS.2017.2785348.
\bibitem{tan2024adaptive}
B.~Tan and J.~Zhao, ``Data-driven adaptive unscented Kalman filter for time-varying inertia and damping estimation of utility-scale IBRs considering current limiter,'' \emph{IEEE Trans. Power Syst.}, vol.~39, no.~6, pp.~7331--7345, 2024, doi: 10.1109/TPWRS.2024.3379956.
\bibitem{Zhao:role}J. Zhao et al., "Roles of dynamic state estimation in power system modeling, monitoring and operation," {\em IEEE Trans. Power Syst.}, vol. 36, no. 3, pp. 2462-2472, May 2021,
\bibitem{Liu:protection} Y. Liu et al., "Dynamic state estimation for power system control and protection," {\em IEEE Trans. Power Syst.}, vol. 36, no. 6, pp. 5909-5921, Nov. 2021.
\bibitem{Qi:Euler}J. Qi, K. Sun, J. Wang and H. Liu, ``Dynamic state estimation for multi-machine power system by unscented Kalman filter with enhanced numerical stability," {\em IEEE Trans. Smart Grid}
\bibitem{Tan:RK}B. Tan, J. Zhao and M. Netto, ``A general decentralized dynamic state estimation with synchronous generator magnetic saturation," {\em IEEE Trans. Power Syst.}, vol. 38, no. 1, pp. 960-963, Jan. 2023.
\bibitem{Lara:revisit}J. D. Lara, R. Henriquez-Auba, D. Ramasubramanian, S. Dhople, D. S. Callaway and S. Sanders, ``Revisiting power systems time-domain simulation methods and models," {\em IEEE Trans. Power Syst.}, 2023.
\bibitem{Katanic:BackEuler}M. Katanic, J. Lygeros, G. Hug. ``Recursive dynamic state estimation for power systems with an incomplete nonlinear DAE model," {\em IET Gener Transm Distrib}, vol. 18, no. 22, pp. 3657-3668, 2024.
\bibitem{Qiu:trapez}J. Qiu, Y. Liu, B. Wang, Y. Xie and W. Huang, ``Protection of multi-terminal hybrid transmission lines based on dynamic states estimation," {\em2023 IEEE PESGM}, Orlando, FL, USA, 2023, pp. 1-5.
\bibitem{Xingyu:GFM}
X.~Zhao, M.~Netto, and J.~Zhao, ``A novel discrete-time state-space model for decentralized dynamic state estimation of grid-forming inverters,'' \emph{IEEE Trans. Power Syst.}, 2025.
\bibitem{Zhao:Robust}X. Zhao and J. Zhao, "A robust and reduced-order power system dynamic state estimator for grid-forming inverters," {\em IEEE Trans. Instrum. Meas.}, 2025.
\bibitem{Singh2014}A. K. Singh and B. C. Pal, ``Decentralized dynamic state estimation in power systems using unscented transformation," {\em IEEE Trans. Power Syst.}, vol. 29, no. 2, pp. 794-804, March 2014.
\bibitem{Zhang:PV}J. Zhang, T. Bi, and H. Liu, ``Dynamic state estimation of a grid-connected converter of a renewable generation system using adaptive cubature Kalman filtering,” {\em Int. J. Electr. Power Energy Syst.}, vol. 143, p. 108470, 2022
\bibitem{Yu:DFIG}S. Yu, T. Fernando, K. Emami and H. H. -C. Iu, ``Dynamic state estimation based control strategy for DFIG wind turbine connected to complex power systems," {\em IEEE Trans. Power Syst.}, vol. 32, no. 2, pp. 1272-1281, March 2017.

\bibitem{Zhu:DFIG} M. Zhu, H. Liu, J. Zhao, B. Tan, T. Bi and S. S. Yu, ``Dynamic state estimation for DFIG with unknown inputs based on cubature Kalman filter and adaptive interpolation," {\em J. Mod. Power Syst. Clean Energy}, vol. 11, no. 4, pp. 1086-1099, July 2023.

\bibitem{Ma:DFIG} W. Ma, C. Wang, L. Dang, X. Zhang and B. Chen, ``Robust dynamic state estimation for DFIG via the generalized maximum correntropy criterion ensemble Kalman Filter," {\em IEEE Trans. Instrum. Meas.}, vol. 72, pp. 1-13, 2023.

\bibitem{Huang:DSE}H. Huang, Y. Lin, X. Lu, Y. Zhao and A. Kumar, ``Dynamic state estimation for inverter-based resources: a control-physics dual estimation framework," {\em IEEE Trans. Power Syst.}, vol. 39, no. 5, pp. 6456-6468, Sept. 2024.

\bibitem{Zhao:GFM} X. Zhao, B. Tan and J. Zhao, ``Power system dynamic state estimation of grid-forming inverters with current limiter," {\em IEEE Trans. Power Syst.}, 2024.
\bibitem{Huang:DSE2} H. Huang and Y. Lin, ``Switching dynamic state estimation and event detection for inverter-based resources with multiple control modes," {\em IEEE Trans. Power Syst.}, 2024.
\bibitem{odunlami2025hybrid}
B.~G.~Odunlami and M.~Netto, ``Dynamic state estimation of hybrid systems: Inverters that switch between grid-following and grid-forming control schemes,'' \emph{arXiv preprint} arXiv:2511.13872, 2025.
\bibitem{Julier2000UT}
S.~Julier, J.~Uhlmann, and H.~F. Durrant-Whyte, ``A new method for the nonlinear transformation of means and covariances in filters and estimators,'' \emph{IEEE Trans. Autom. Control}, vol.~45, no.~3, pp.~477--482, Mar.~2000.
\bibitem{Sarkka2013}
S.~S{\"a}rkk{\"a},
\emph{Bayesian Filtering and Smoothing},
Institute of Mathematical Statistics Textbooks, no.~3.
Cambridge, U.K.: Cambridge Univ. Press, 2013.
\bibitem{Zhao:GFL:Note}
X.~Zhao, M.~Netto, and J.~Zhao, ``Input-Output State-Space Representation of Grid-Following Inverters,'' [Online]. Available: \url{https://drive.google.com/file/d/1SLX5UreMmAeWdhNQysbgBvrl9GdrQVuC/view?usp=sharing}
\bibitem{IEEE Std 60255} ``IEEE/IEC International Standard - Measuring relays and protection equipment - Part 118-1: Synchrophasor for power systems - Measurements," {\em IEC/IEEE 60255-118-1:2018}, pp. 1-78, 2018.


\end{thebibliography}
\end{document}